\documentclass[10pt,preprint]{aastex}

\slugcomment{Scheduled to appear in the Astronomical Journal, Vol 127, n.6, 2004 June}
\shorttitle{The nature of the mid-infrared population} 
\shortauthors{La Franca, Gruppioni, Matute et al.}

\begin{document}

   \title{The nature of the mid-infrared population from optical identifications of the
ELAIS-S1 sample}

\author{F. La Franca\altaffilmark{1}, 
C. Gruppioni\altaffilmark{2},
I. Matute\altaffilmark{1},
F. Pozzi\altaffilmark{2,3},
C. Lari\altaffilmark{4},
M. Mignoli\altaffilmark{2}, 
G. Zamorani\altaffilmark{2}, 
D.M. Alexander\altaffilmark{5}, 
F. Cocchia\altaffilmark{6},
L. Danese\altaffilmark{7}, 
A. Franceschini\altaffilmark{8},
P. H\'{e}raudeau\altaffilmark{9}, 
J.K. Kotilainen\altaffilmark{10},
M.J.D. Linden-V\o rnle\altaffilmark{11},
S. Oliver\altaffilmark{12}, 
M. Rowan-Robinson\altaffilmark{13},
S. Serjeant\altaffilmark{14}, 
L. Spinoglio\altaffilmark{15} 
\& A. Verma\altaffilmark{16}
}

\altaffiltext{1}{Dipartimento di Fisica, Universit\`a degli Studi "Roma Tre",
Via della Vasca Navale 84, I-00146 Roma, Italy.}
\altaffiltext{2}{INAF, Osservatorio Astronomico di Bologna, Via Ranzani 1, I-40127 Bologna, Italy}
\altaffiltext{3}{Dipartimento di Astronomia, Universit\`a di Bologna, Via Ranzani 1, I-40127 Bologna, Italy}
\altaffiltext{4}{INAF-CNR, Istituto di RadioAstronomia (IRA), via Gobetti 101, I-40129, Bologna, Italy}
\altaffiltext{5}{Institute of Astronomy, Madingley Road, Cambridge, CB3 OHA, UK}
\altaffiltext{6}{INAF, Osservatorio Astronomico di Roma, Via Frascati 33, Monteporzio-Catone, I-00040, Italy}
\altaffiltext{7}{SISSA, International School for Advanced Studies, Via Beirut 2-4, 34014 Trieste, Italy}
\altaffiltext{8}{Dipartimento di Astronomia, Universit\`a di Padova, vicolo dell'Osservatorio 5,
 I-35122, Padova, Italy}

\altaffiltext{9}{Kapteyn Astronomical Institute, Landleven 12, 9747 AD Groeningen, The Netherlands}
\altaffiltext{10}{Tuorla Observatory, University of Turku, V\"{a}is\"{a}l\"{a}ntie 20, FIN-21500, Piikki\"{o}, Finland}
\altaffiltext{11}{Niels Bohr Institute for Astronomy, Physics and Geophysics, Astronomical Observatory,
Juliane Maries Vej 30, DK-2100 Copenaghen, Denmark}
\altaffiltext{12}{Astronomy Centre, Department of Physics \& Astronomy, University of Sussex, Brighton,
BN1 9QJ, UK}
\altaffiltext{13}{Astrophysics Group, Blackett Laboratory, Imperial College of Science and Technology
\& Medicine (ICSTM), Prince Consort Rd., London, SW7 2BZ, UK}
\altaffiltext{14}{Centre for Astrophysics and Planetary Sciences, School of Physical Sciences,
University of Kent, Canterbury, Kent CT2 7NR, UK}
\altaffiltext{15}{INAF-CNR, Istituto di Fisica dello Spazio Interplanetario, Via del Fosso del Cavaliere 100, 00133 Roma, Italy}
\altaffiltext{16}{Max-Planck-Institut f\"{u}r extraterrestrische Physics, Postfach 1603, 85740 Garching, Germany}

\begin{abstract}
We present a multi-wavelength catalog (15 $\micron$, R, K-band, 1.4
GHz flux) plus spectroscopic identifications for 406 15 $\micron$
sources detected in the European Large Area {\it ISO} Survey (ELAIS)
region S1, over the flux density range 0.5$<$$S_{15\micron}$$<$150
mJy. 332 ($\sim$82\%) sources are optically identified down to
R$\sim$23.0. Spectra or bona fide stellar identifications are obtained
for 290 objects ($\sim$88\% of the optically identified sources). The
areal coverage, mid-infrared (MIR) and optical completeness of the
sample are discussed in order to allow statistical and evolutionary
analyses.  Two main spectroscopic classes have been found to dominate
the MIR extragalactic population: $z<0.5$ star-forming galaxies (from
absorbed to extreme starbursts: $\nu L_{\nu}(15 \micron)$$\approx
10^{8}-10^{11}$~L$_{\odot}$), which account for $\sim$75\% of the
sources, and Active Galactic Nuclei (AGN; both type 1 and 2), which
account for $\sim$25\% of the sources. About 20\% of the extragalactic
sources are dust-enshrouded starburst galaxies [e(a) spectra], and all
the starburst galaxies appear more dust extincted in the optical than
nearby normal galaxies. We also identified 91 stellar objects
($\sim$22\% of the MIR sources). The counts for starburst galaxies and
AGN down to 0.6 mJy have been derived. A general trend is found in the
optical-MIR spectral energy distribution (SED) of the galaxies, where
the MIR-luminous objects have larger MIR to optical luminosity
ratios. Based on a variety of analyses, we suggest that the ELAIS
sources fainter than R$\sim$23 are luminous and ultra-luminous MIR
galaxies (LIG-ULIGs; $\nu L_{\nu}(15 \micron)$=10$^{11}$-10$^{12}$
L$_\odot$) at intermediate redshifts ($z$=0.5--1.5), and that
consequently the present sample is virtually 100\% spectroscopically
complete up to $z$=0.5.

\end{abstract}


\keywords{
Cosmology: observations; Galaxies: distances and
redshift; Galaxies: evolution; Galaxies: active; Galaxies: starburst; Infrared: galaxies
}

\section{Introduction}\label{intro}

The ISOCAM instrument (Cesarsky et al. 1996) on board of the {\it
Infrared Space Observatory} ({\it ISO}; Kessler et al. 1996) has been
able to unveil at 15 $\micron$ a population of MIR galaxies with two
orders of magnitude fainter flux densities than observed by the {\it
Infrared Astronomical Satellite} ({\it IRAS}).  These galaxies result to be
ten times more numerous than expected if there were no evolution from
$z = 0$ up to $z= 1 - 1.5$ (see Elbaz et al. 1999). 
These results are confirmed by the detection of a substantial diffuse
cosmic Infrared Background in the 140 $\mu$m - 1 mm range (Puget et
al. 1996; Hauser et al.  1998; Fixsen et al. 1998; Lagache et
al. 1999), which suggests a strong evolution of the galaxies
in the IR (stronger than observed at any other wavelengths). The
suggested evolution can either imply a larger fraction of galaxies
experiencing an IR luminous phase in the past, or that MIR galaxies
were more luminous in the past (see e.g. Chary \& Elbaz 2001).

So far, the nature of the sources responsible for the strong evolution
observed in the MIR band has only been studied in small fields,
containing sources well known at other wavelengths: the Hubble Deep
Field North (HDF-N; Aussel et al. 1999a,b) and South (HDF-S; Oliver et
al. 2002; Mann et al. 2002; Franceschini et al. 2003), the
Canada-France Redshift Survey (CFRS) 1452+52 field (Flores et
al. 1999) and the ELAIS-S2 field (Pozzi et al. 2003). This has yielded
a small but meaningful sample of sources sufficient for most
multiwavelength studies. Most of these sources can easily be
identified with relatively bright optical counterparts ($I < 22.5$)
with a median redshift of $z\simeq0.7-0.8$ imposed by the MIR {\it
k}-correction (Aussel et al. 1999a,b; Flores et al. 1999).

However, to obtain reliable source counts and statistically
significant information about the nature and evolution of the MIR
extragalactic source population, large-area surveys covering a wide
range of flux densities are required. In particular, it is very
important to bridge the gap between the {\it IRAS} surveys
($S_{15\micron}>200$ mJy) and the Deep ISOCAM surveys ($0.1<
S_{15\micron}<2-3$ mJy). ELAIS (Oliver et~al. 2000), with its broad
flux range (0.5$<S_{15\micron}<$150 mJy) and several thousand sources
detected over a $\sim$12 deg$^2$ area (Rowan-Robinson et~al. 2004), is
the crucial survey to investigate the nature and evolution of the MIR
sources in the region where the source counts start diverging from
the non-evolution predictions (a few mJy at 15 $\mu$m; Elbaz et
al. 1999; Gruppioni et al. 2002) and with data not affected by 
cosmic variance (which might be a strong effect in the narrow area 
surveys).

In order to better understand the nature of the MIR extragalactic
population on the basis of a larger identification statistics than the
ones obtained so far, we performed photometric and spectroscopic
identifications for a source sample selected at 15 $\mu$m in the ELAIS
southern field S1 ($\sim$4 deg$^2$). Here we present the results of
the identification of 406 objects, which form a complete and highly
reliable sub-sample drawn from the 15 $\micron$ catalog of Lari
et~al. (2001). The areal coverage, MIR and optical completeness are
discussed in detail to allow statistical and evolutionary analyses
with this sample. The counts down to 0.6~mJy for each class of source
have been derived. The multiwavelength (radio, MIR, near-infrared
(NIR), optical) properties of the 15 $\mu$m sources are investigated
in order to study the nature and evolution of the infrared
extragalactic population. The infrared luminosity function and the
evolution of 15 $\mu$m galaxies and AGN are discussed in two
companion papers (Pozzi et al. 2004; Matute et al. 2004, in
preparation). A preliminary analysis of the AGN evolution has been
presented in Matute et~al. (2002) \footnote{A collection of papers and 
scientific results related to the ELAIS southern fields is available
at \url{http://www.fis.uniroma3.it/$\sim$ELAIS\_S}. }


This paper is structured as follows: in section 2 we describe the 15
$\micron$ sample; in section 3 we present the optical CCD
observations, the source identification and the cross-correlation with
catalogs in other bands; in section 4 we describe the spectroscopic
observations, the classification and the spectroscopic catalog; in
section 5 we present the counts; in section 6 we discuss the nature of
the MIR galaxies.

Throughout this paper we will assume $H_0 = 75$ km s$^{-1}$ Mpc$^{-1}$, 
$\Omega_m = 0.3$ and $\Omega_{\Lambda} = 0.7$.


\section{The ELAIS-S1 Mid-Infrared Sample}

%
%

The primary {\it ISO} survey in the ELAIS regions was restricted to
two bands: 15 $\mu$m (using ISOCAM), and 90 $\mu$m (using ISOPHOT;
Lemke et al. 1996). Additional observations were performed in
some parts of the survey area at 6.7 $\mu$m and 175 $\mu$m.  The main
survey area, consisting of $\sim$12 deg$^2$, is divided into four
fields distributed across the sky in order to decrease the biases due
to cosmic variance: three in the northern hemisphere (N1, N2 and N3)
and one in the southern hemisphere (S1). Details of the survey can be
found in Oliver et al. (2000). The final band merged catalog (6.7, 15,
90 and 175 $\micron$) of the whole ELAIS survey is in Rowan-Robinson
et~al. (2003)\footnote{See
\url{http://astro.imperial.ac.uk/elais/}.}.

The southern field, S1, centered at $\alpha$(2000) = 00$^h$ 34$^m$
44.4$^s$, $\delta$(2000) = -43$^{\circ}$ 28$^{\prime}$ 12$^{\prime
\prime}$, covers an area of the sky of about $2^{\circ}$$\times$$2^{\circ}$.
In this field, a new method of data reduction ({\it LARI
technique}), especially developed for the detection of faint sources,
has been successfully applied, producing a catalog (complete at the
5$\sigma$ level) of 462 sources in the flux density range 0.5 -- 150
mJy\footnote{After flux corrections as explained at the end of \S 2.}. 
Details about the data reduction technique and the source
catalog 
have been presented in Lari et al. (2001)\footnote{The catalog is available at
\url{http://www.bo.astro.it/$\sim$elais/catalogs/ELAIS\_CAM\_15micron\_S1.TAB}.}, 
while the sample completeness
and the source counts at 15 $\mu$m obtained from that catalog have
been presented and discussed by Gruppioni et al. (2002).


The whole S1 area has been surveyed in the radio with the Australia
Telescope Compact Array down to $S_{1.4 GHz} \simeq$ 0.4~mJy (5$\sigma$,
Gruppioni et al. 1999)\footnote{ Available at
\url{http://www.bo.astro.it/$\sim$elais/catalogs/ELAIS\_RADIO\_S1.TAB}.}.
The radio catalog in S1 consists of 652
sources detected at the $5\sigma$ level at 1.4 GHz.  The radio-MIR
correlation for starburst galaxies has been discussed by Gruppioni et
al. (2003). About 40\% the area has been observed by {\it BeppoSAX}
down to a 2-10 keV sensitivity of $\sim$10$^{-13}$ erg s$^{-1}$ cm$^{-2}$ by
Alexander et al. (2001).

\begin{figure}
\epsscale{0.7}
\plotone{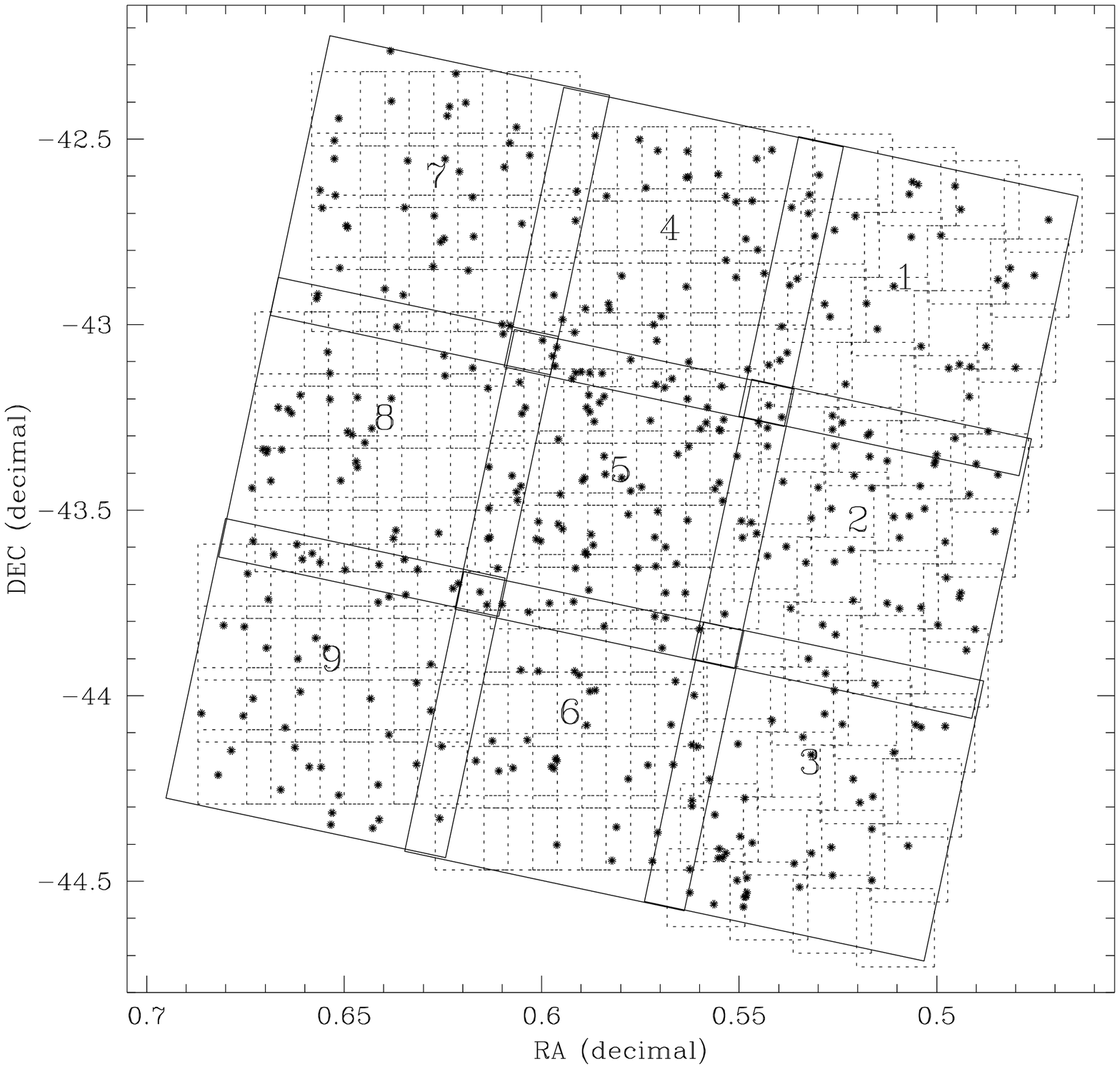}
\caption{
ELAIS-S1 field divided into the 9 ISOCAM rasters. The total area is 4.2
deg$^2$. The numbers refer to the individual rasters.  The 168 CCD
pointings and the 406 ELAIS-S1 sources are shown by dotted lines and
filled points, respectively.  }
\label{field}
\end{figure}

\begin{figure}
\resizebox{\hsize}{!}
{\includegraphics[angle=-90]{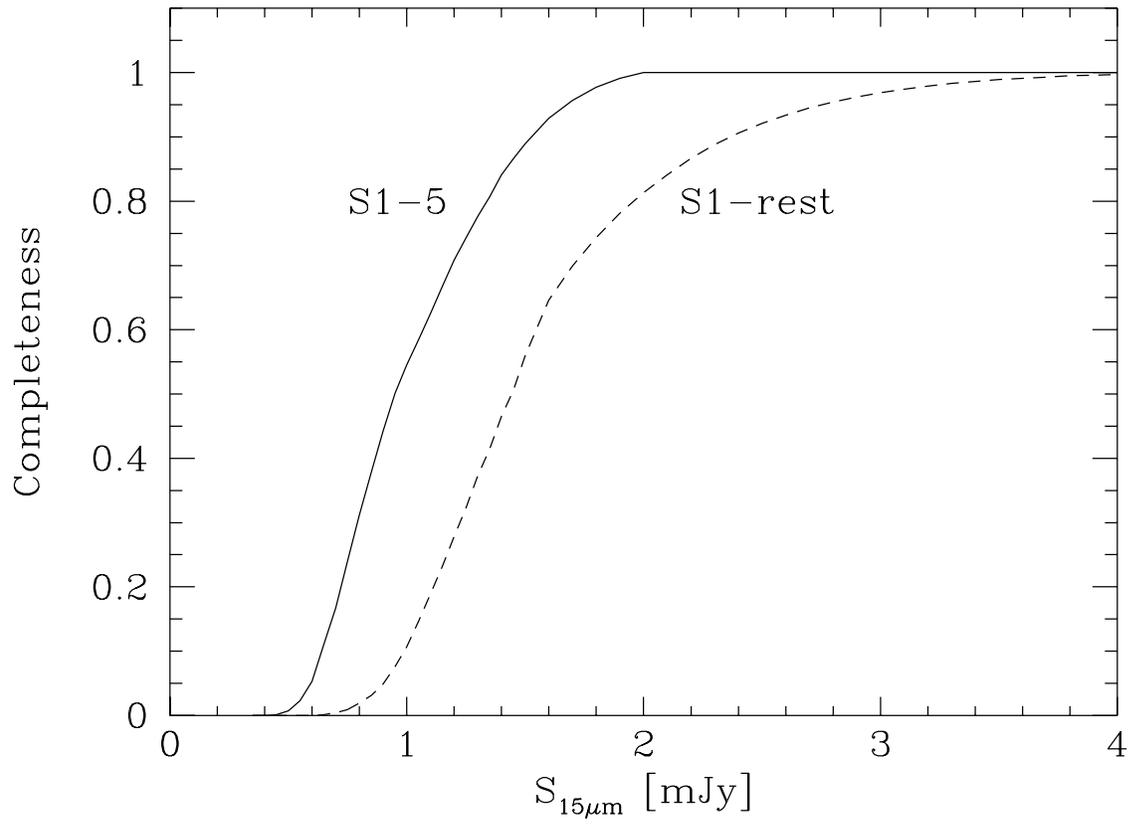}}
\caption{
Completeness as a function of the 15 $\micron$ flux of the S1-5 (continuous line)
and the S1-rest (dashed line) regions.
 }
\label{fig_compl}
\end{figure}

In this work we restricted our analysis to a highly reliable subsample
of 406 sources out of a total 462 15 $\mu$m sources. The ELAIS-S1 area
has been observed by ISOCAM at 15 $\mu$m with 9 separate rasters (see
Figure \ref{field}). 20 out of the 462 sources were excluded as they
were detected twice at the edges of overlapping rasters (in this case
the sources with the highest S/N ratio have been selected). Moreover,
36 sources were excluded as we decided to use only those sources with
the highest reliability (REL=0) after a visual inspection of their
pixel histories. The ISOCAM observations of the central (S1\_5) raster
have been repeated three times (instead of one) and thus in this part
the survey reaches fainter fluxes. As a consequence, we have divided
the sample into two regions: a) S1\_5 (i.e. raster 5) which covers
0.55 deg$^2$ and reaches a 20$\%$ completeness at $\sim$0.7 mJy, and
b) S1\_rest (i.e. all the remaining rasters) which covers 3.65 deg$^2$
and reaches a 20$\%$ completeness at $\sim$1.1 mJy.  Note that in this
work, as well as in the counts paper (Gruppioni et al. 2002), we used
the 15 $\micron$ fluxes of the ELAIS-S1 sources reported by Lari et
al. (2001), corrected following the detailed recipe described by
Gruppioni et al. (2002). The completeness functions of the two regions
as a function of the corrected 15 $\micron$ flux are shown in Figure
\ref{fig_compl} and are reported in Table \ref{tab_compl}, while the
corrected 15 $\micron$ fluxes are reported in Table \ref{tab_id} and
Table \ref{tab_lines}.


\section{Optical photometry and multi-wavelength identifications}
\subsection{CCD Observations and Data Reduction}

In order to obtain R-band optical photometry of the ELAIS-S1 region
down to magnitudes fainter than the typical Schmidt plates limits
(e.g. COSMOS\footnote{\url{http://xip.nrl.navy.mil/www\_rsearch/RS\_form.html}.}),
we have carried out a CCD imaging campaign at the European Southern
Observatory (ESO\footnote{\url{http://www.eso.org}.}). The whole S1
region has been covered by 168 overlapping CCD exposures obtained with
DFOSC mounted on the 1.54 Danish/ESO telescope at La Silla
(Chile). The observing runs were carried out during 12 nights (almost
all photometric) during September 1996, October 1996 and September
1997. The CCD had 2052$\times$2052 pixel array, with a pixel scale of
0.39$''$, and a field of view of 13.3$\times$13.3 arc-minutes.  The
filter used was the Bessel R (ESO
\#452). About 97\% the ELAIS-S1 area has been observed in the R-band (see Figure
\ref{field}).

Photometric calibration was carried out by frequently observing
(every $\sim$1-1.5 hours) the standard field stars ``T Phe'' ($\alpha$(2000)
= 00$^h$ 27$^m$ 49.0$^s$, $\delta$(2000) = -46$^{\circ}$ 48$^{\prime}$
02$^{\prime
\prime}$; Landolt 1992). Those fields (about 10$\%$) observed without
photometric conditions
were later calibrated using shorter integration images taken during
photometric weather.  Flat fields were obtained from blank fields in
the sky at sunset and sunrise every night. The integration time for all
science frames was 20 minutes.  Standard data reduction was carried
out with the MIDAS and IRAF reduction packages. Every science frame
was subtracted for bias, and corrected for flat field.

Source searching on the final reduced images was carried out with
SExtractor (Bertin \& Arnouts 1996) The ``MAG\_BEST'' source
magnitudes have been used. The R-band magnitudes were computed after
correction for airmass extinction, and without a color term,
i.e. assuming a $V-R=0$ color for all the objects.  Due to the long
observing campaign the characteristics of the CCD observations
(e.g. seeing and magnitude limits) varied from field to field.  The
magnitude limits vary from 22.5 to 23.0 at 95\% completeness
level. 
The astrometry has been computed by using the USNO2 catalog\footnote{
\url{http://www.nofs.navy.mil/}.} as reference, with an internal
accuracy of 0.26$''$ (rms), and an absolute radial accuracy (compared
to USNO2) of 0.38$''$ (rms).

\subsection{Identification of the 15 $\mu$m Sources in the $R$ band}

\begin{figure}
\plotone{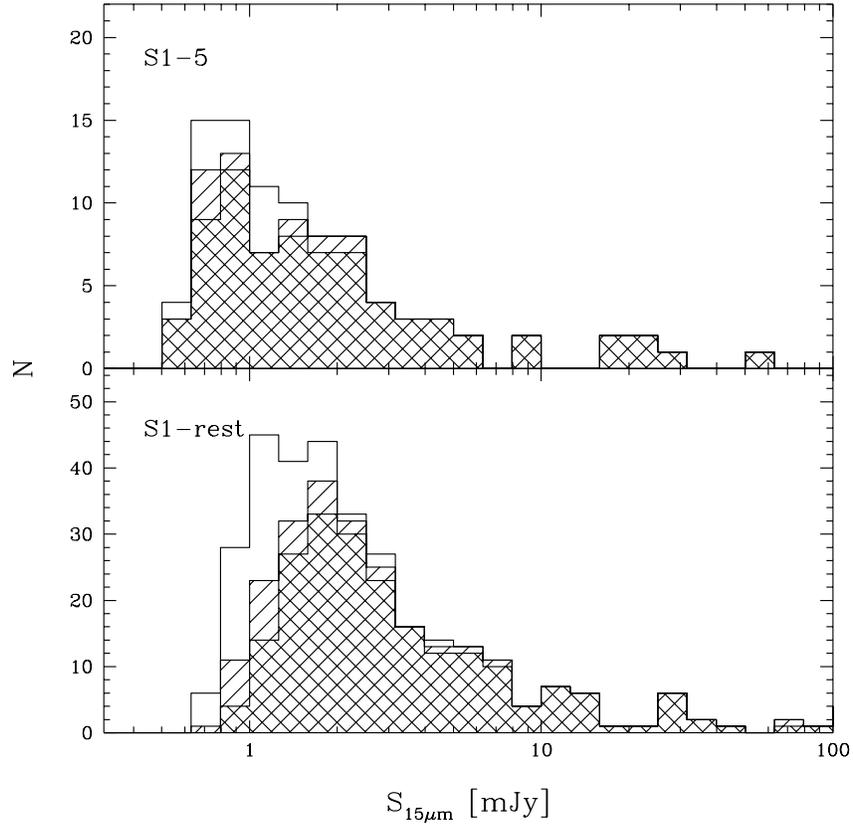}
\caption{
15$\mu$m flux histograms of the 406 ELAIS-S1 sources
in the S1-5 ($Top$) and S1-rest ($Bottom$) regions.  The
shaded histograms are the optically identified sources, and the
cross-shaded histograms are the spectroscopically identified sources.
\label{fig_histF}
}

\end{figure}

For the optical identifications of the ISOCAM sources we used the $R$
magnitudes and positions given in the CCD catalog. Ten sources were
not observed by the CCD imaging. For 7 out of these 10 sources the
R-band USNO2 magnitudes have been used. The remaining 3 sources
(ELAISC15\_J003559-430232, ELAISC15\_J003912-442051,
ELAISC15\_J004023-433459) have to be considered fainter than R=20 (see
Table \ref{tab_id}). Two sources (ELAISC15\_J003034-433428 and
ELAISC15\_J003857-431719) were too close to bright stars and the
optical identification was not possible. To estimate the
identification reliability, we adopted the likelihood ratio technique
as described by Sutherland \& Saunders (1992) (see Ciliegi et
al. (2003) for a detailed description). When possible counterparts
fainter than R=23.0 were found, the identifications were not
statistically discussed, and presented only as tentative.  Taking into
account all of the optical identifications, 332 ($\sim$82\%) out of
the 406 ELAIS-S1 sources have a reliable counterpart down to R$\sim$23
(only 7 sources have 23.0$<$R$<$24.3), 2 are impossible to identify,
and 72 ($\sim$18$\%$) are to be considered fainter than R$\sim$23.0
(see Figure
\ref{fig_histF}). In the following we will call these sources
``EMPTY''.


\subsection{K band and 1.4 GHz identifications}

The sample has been cross-correlated with the K-band catalog from the
Two Micron All Sky Survey\footnote{See
\url{http://www.ipac.caltech.edu/2mass}.} (2MASS),
which reaches a K-band limit of
$\sim$15.6 (at 95\% completeness
level). In total 211 sources have been detected in the K-band.
Since the whole S1 field has been surveyed in the radio (1.4 GHz) with
the Australia Telescope Compact Array to an average 1$\sigma$ level of
0.08 mJy (Gruppioni et al. 1999), at each ISOCAM position, we have
searched for detection in the radio map down to 3 $\sigma$, finding 71
likely radio counterparts (61 with measured redshifts; see Gruppioni et
al. 2003 for details). The results of the multi-band identifications
are reported in Table \ref{tab_id}.

\section{The spectroscopic catalog}

\subsection{Spectroscopic observations}

Spectroscopic observations of the optical counterparts of the ISOCAM
S1 sources were carried out at the 2dF/AAT and ESO Danish 1.5-m, 3.6-m
and NTT telescopes during September and October 1998, September and
November 2001, September and October 2002. A total of 15 ESO nights
and 2 UK (from the Panel for the Allocation of Telescope Time; PATT)
nights were allocated.

The 2dF/AAT multifiber feed spectrograph was used with 200 fibers
equipped with the 300B grism and the remaining 200 fibers with the
270R grism (about 10 \AA\ instrumental resolution). Because of poor
weather conditions only half an hour of integration was possible during
each night,
reaching a magnitude limit of R$\sim$19.5. The ESO spectrographs
(DFOSC, EFOSC2 and EMMI) were all used in single slit low resolution
mode (10-20 \AA\ resolution) in the wavelength range 3500-9500 \AA,
and reaching an average S/N ratio per resolution element of
$\sim$15-30 in the wavelength range 4000-7000 \AA.  The
reduction process used standard MIDAS and IRAF facilities.  The raw
data were sky-subtracted and corrected for pixel-to-pixel variations
by division with a suitably normalized exposure of the spectrum of an
incandescent source (flat field). Wavelength calibrations were carried
out by comparison with exposures of He, Ar and Ne lamps. Relative flux
calibration was carried out by observations of spectrophotometric
standard stars (Oke 1990, Hamuy 1992, 1994).


\subsection{Classification}

We have spectroscopically classified 290 ($\sim$87\%) out of the 332
sources optically identified ($\sim$71\% of the whole 15 $\mu$m
sample). The sample has been cross-correlated with the TYCHO2 (H\o g
et al. 2000) and NED\footnote{See
\url{http://nedwww.ipac.caltech.edu/}.} catalogs in order to identify
known stars and extragalactic objects. Sixty-two TYCHO2 stars were
found. The remaining objects brighter than R=15 were visually
inspected in the R band images and those suspected to be
extended/extragalactic were spectroscopically observed. Fifteen
sources with R$<$15 were considered to be bona fide stars on the basis
of the visual inspection, while 14 were spectroscopically identified
as stars at the telescope. In total we found 91 stars (all of them
brighter than R=15) and 199 extragalactic sources.

In order to classify the extragalactic sources we have first
identified type 1 AGN (AGN1) when permitted lines with rest frame full
width half maximum (FWHM) larger than 1200 Km/s were found.  The
remaining objects were classified according to two methods.

\begin{figure}
\plotone{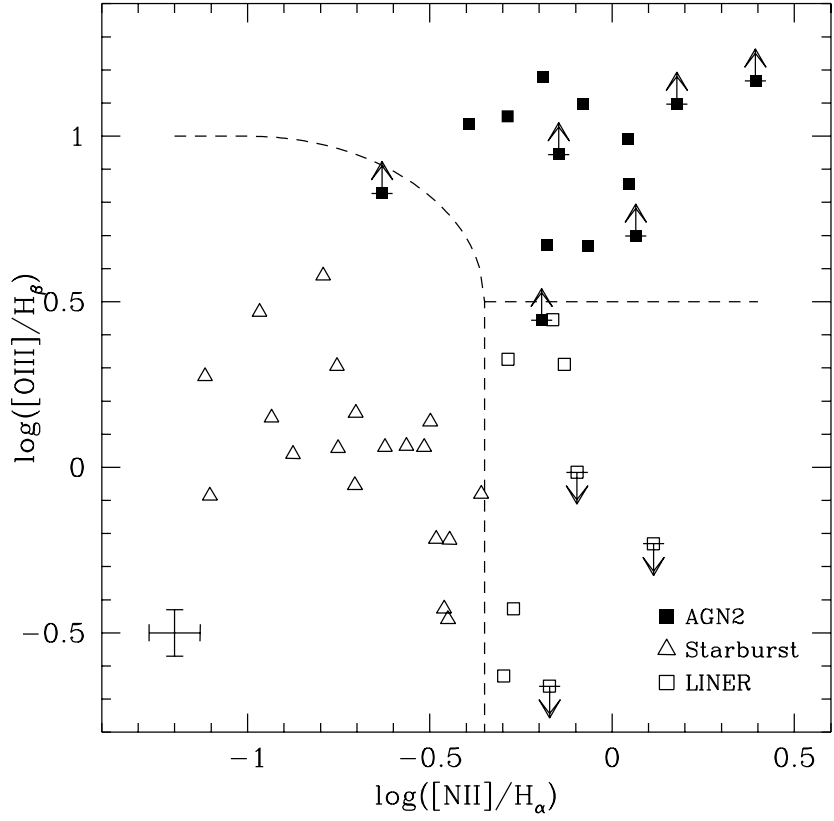}
\caption{Diagnostic diagram for the classification of emission line galaxies.
The dashed lines separate the regions of AGN2, LINERs and starburst galaxies
(see e.g. Veilleux \& Osterbrock 1987 and Tresse et al. 1996). The errorbars
correspond to 10\% uncertainties.}
\label{fig_diagn}
\end{figure}

The first method was used to identify type 2 AGN (AGN2), and to
separate them from other emission line galaxies such as HII/starburst
and Low Ionization Narrow Emission Regions (LINER).  All the objects
with narrow H$\beta$ or [OIII]$\lambda$5007 emission lines having EW
(measured as the ratio between the integrated flux in the line, as
resulting from a gaussian fit, and the flux of the continuum at the
line wavelength) larger than $\sim$3-5\AA\ (our line detection limit),
were classified according to three diagnostic diagrams: 1)
[OIII]$\lambda$5007/H$\beta$ ratio versus [NII]$\lambda$6583/H$\alpha$
ratio; 2) [OIII]$\lambda$5007/H$\beta$ ratio versus
[SII]$\lambda$6725/H$\alpha$ ratio; 3) [OIII]$\lambda$5007/H$\beta$
ratio versus [OII]$\lambda$3727/H$\beta$ ratio ( e.g. Veilleux \&
Osterbrock 1987; Tresse et al. 1996; see Figure \ref{fig_diagn} for an
example). The H$\alpha$ and [NII]$\lambda$6583 lines were deblended
using the IRAF task {\tt splot}, which assumes Gaussian line profiles
and a linear background.

The above method is efficient in separating AGN from starburst
activity, but does not provide information on the nature of a large
fraction of our galaxies, which do not show [OIII] and H$\beta$
emission lines. We have thus applied the classification scheme of
Dressler et al. (1999) (see also Poggianti et al. 1999 and Poggianti
\& Wu 2000) to all the ``non AGN'' galaxies. This method is based
primarily on two lines, [OII]$\lambda$3727 in emission and H$\delta$
in absorption, which are good indicators of (respectively) current and
recent star formation (SF) in distant galaxy spectra. Under this
scheme the spectra have been divided in:

\begin{enumerate}
\item{} e(a) (having 3\AA$<$EW([OII])$<$40\AA\ and
EW(H$\delta$)$\geq$4\AA), spectra of dust-enshrouded starburst
galaxies;

\item{} e(b) (having EW([OII])$>$40\AA), spectra with very strong emission
lines of galaxies which are undergoing strong star formation;

\item{} e(c) (having 3\AA$<$EW([OII])$<$40\AA\ and EW(H$\delta$)$<$4\AA),
typical spectra of spirals which have been forming stars in a
continuous fashion;

\item{} k (absent [OII] and EW(H$\delta$)$<$3\AA), spectra of passive
elliptical-like galaxies with neither ongoing nor recent star formation;

\item{} k+a/a+k (absent [OII] and EW(H$\delta$)$\geq$3\AA), spectra of
poststarburst galaxies with no current star formation which were
forming stars at a vigorous rate in the recent past (last 1.5 Gyr).

\item{} k(e), spectra similar to the k-type but with signs of at least one
emission line. This class was introduced by us following Duc et
al. (2002).

\end{enumerate}



\begin{figure}
\plotone{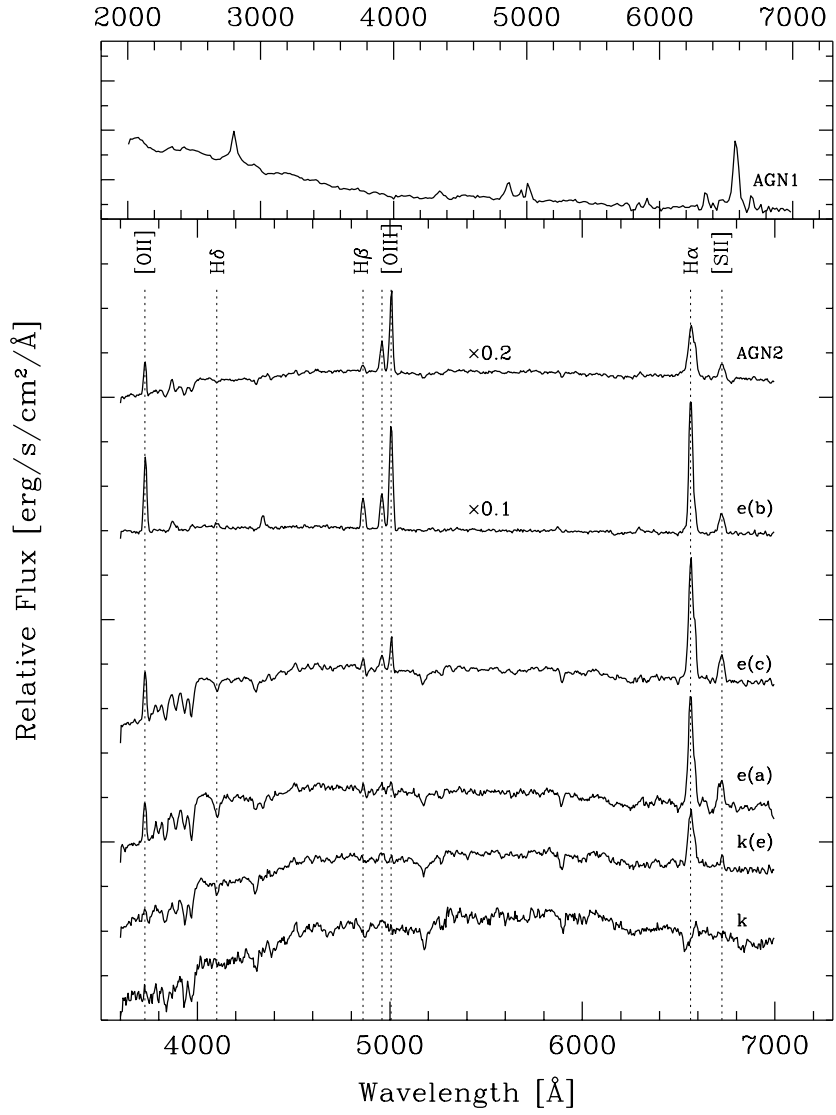}
\caption{
Composite spectra of the various type of galaxies classified
in the ELAIS-S1 survey (see text).}
\label{fig_spettri}
\end{figure}

Following this classification, we found 25 AGN1, 23 AGN2, 37 e(a), 11
e(b), 68 e(c), 32 k(e) and 3 k galaxies. No k+a/a+k galaxy was
found. Thirteen objects could not be classified on the basis of this
scheme because the wavelength range of [OII] and H$\delta$ lines was
not observed (they were observed with the red grism of the 2dF
spectrograph). Two of them were classified e(b) (starburst) galaxies
according to the [OIII]$\lambda$5007/H$\beta$ and
[NII]$\lambda$6583/H$\alpha$ ratios, while the remaining 11 galaxies
were classified k(e) because no [OIII]$\lambda$5007 and H$\beta$ line
was detected, while the H$\alpha$ emission line was present. In Figure
\ref{fig_spettri} we show the composite spectra of each type of galaxy found
under this classification scheme.

We should stress that the number of identified AGN2 has to be
considered a lower limit. {\it XMM} and {\it Chandra} surveys have
revealed significant numbers of (mostly absorbed) AGN with L(2-10
keV)$>$ 10$^{42}$ erg s$^{-1}$ (i.e. AGN2) whose optical spectra do
not show sign of AGN activity (see e.g. Fiore et al.  2000;
2003). Recent (June-July 2003), 100 Ksec long, {\it XMM} observations
of S1-5 have revealed indeed that about 10-20\% of the counterparts of
our {\it ISO} sources classified as no-AGN galaxies on the basis of
their optical spectra, can harbor an X-ray emitting AGN ( the {\it
BeppoSAX} observations presented in Alexander et~al. (2001) are too
shallow to provide tighter AGN constraints).  Under the reasonable
assumption that all these objects are AGN2, the number of AGN2 in the
ELAIS-S1 sample should be increased by at least a factor of 2. A more
detailed discussion on the X-ray/MIR relationship of AGN and starburst
galaxies is beyond the scope of this paper and will be presented by La
Franca et al. (in preparation).



In Table \ref{tab_id} we list the multiband associations to the 15
$\micron$ sources with the likelihood and reliability of the R band
identifications, plus the redshift, the spectroscopic classification
and the $\nu L_\nu$ luminosities at 15 $\micron$ and in the R-band. In
Table
\ref{tab_lines} we present the line measurements of the spectra which
include the calcium break (D$_{4000}$; measured from the fluxes in the
intervals 3750-3950 \AA\ and 4050-4250 \AA; see Dressler et al. 1999)
and the EW for the [OII]$\lambda$3727, H$\delta$, H$\beta$,
[OIII]$\lambda 5007$, H$\alpha$, [NII]$\lambda$6583 and
[SII]$\lambda$6725 lines (positive values correspond to absorption
lines).

\begin{figure}
\centering
\resizebox{\hsize}{!}
{\includegraphics[angle=-90]{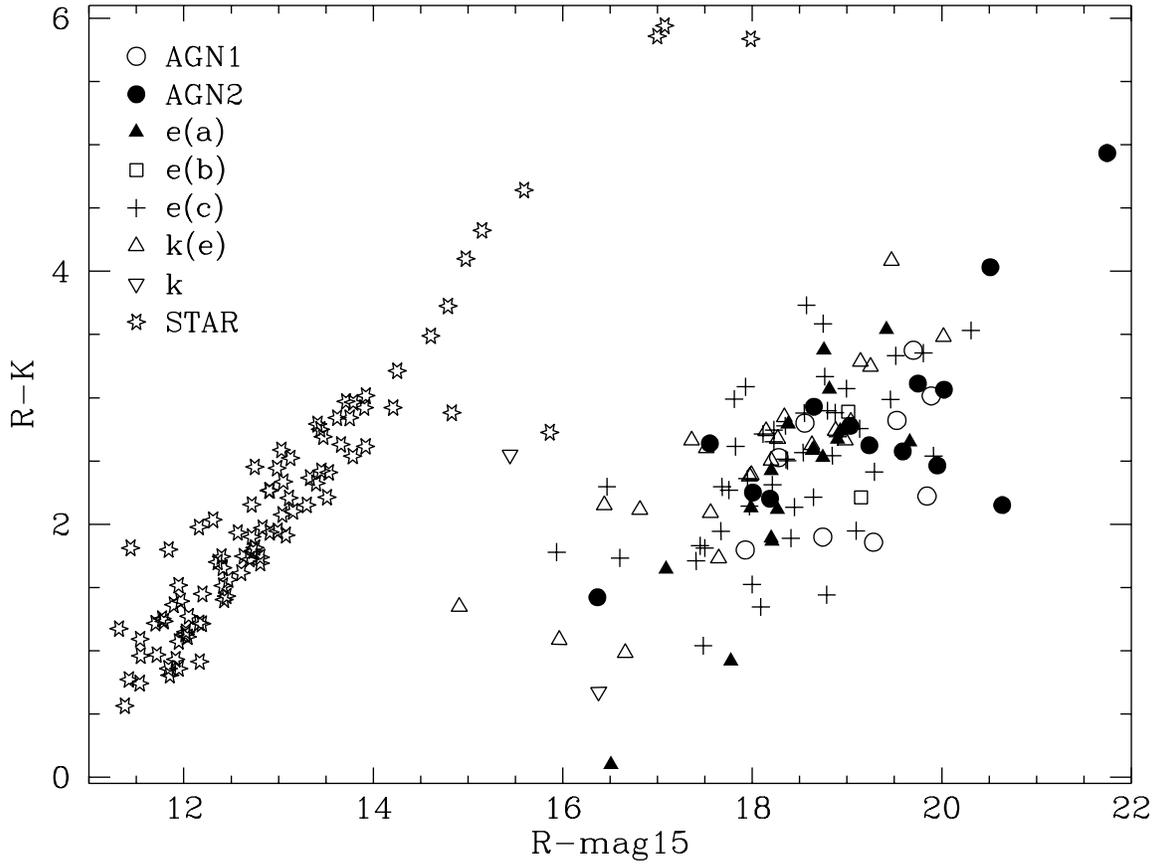}}
\caption{
$R-K$ color versus $R-mag15$ color diagram of the 211 ELAIS-S1 sources
detected both in the R and K bands. $mag15$ is a 15 $\micron$ magnitude
computed as $mag15=-2.5 log(S_{15}(mJy))$.}
\label{fig_R15_RK}
\end{figure}

Figure \ref{fig_R15_RK} shows the distribution of the 211 ELAIS-S1
sources detected both in the R and K bands, in the $R-K$ versus
$R-mag15$ plane, where we have defined $mag15=-2.5
log(S_{15}(mJy))$. This diagram shows a clear separation of the locus
of the stars from the locus of the galaxies: for any given $R-K$
color, stars populate a region with a two order of magnitude lower
MIR to optical luminosity ratio. Note that only one early-type galaxy
and two stars are located in the intermediate area of the diagram
between those occupied by stars and galaxies. This means that stars
can easily be separated from galaxies based purely on their infrared
to optical flux ratio (see also V\"ais\"anen et al. 2002).


\subsection{Completeness}

\begin{figure}
\plotone{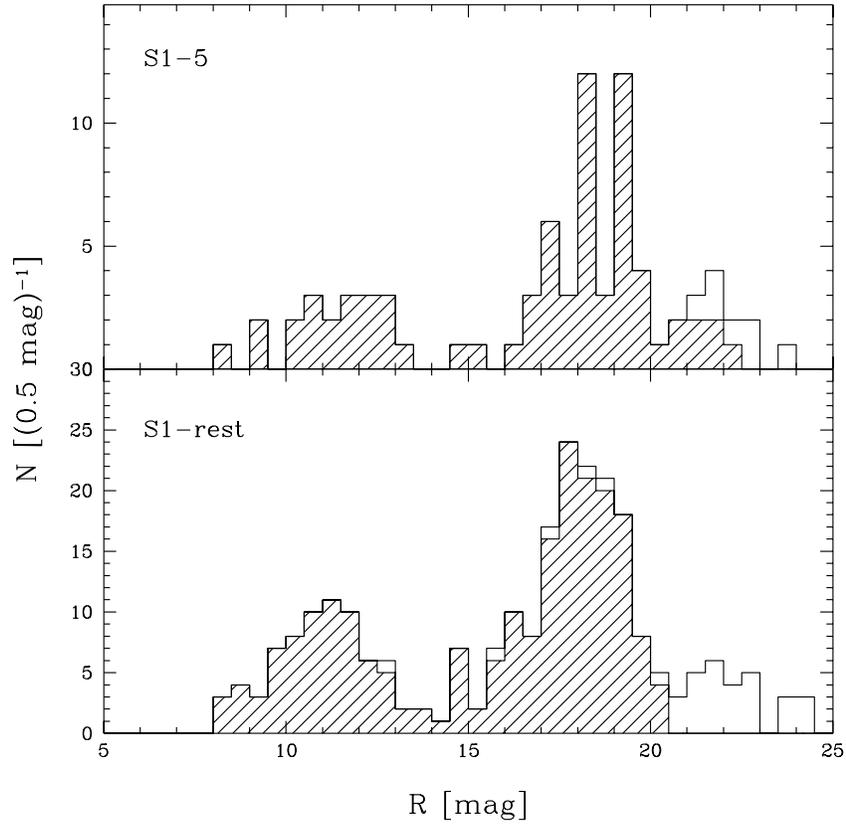}
\caption{
R-band magnitude histograms of the optically identified 15 $\micron$
sources in the S1-rest ($Top$) and S1-5 ($Bottom$) regions.  The
shaded histograms are the R-band distributions of the
spectroscopically identified sources.}
\label{fig_histR}
\end{figure}

\begin{figure}
\plotone{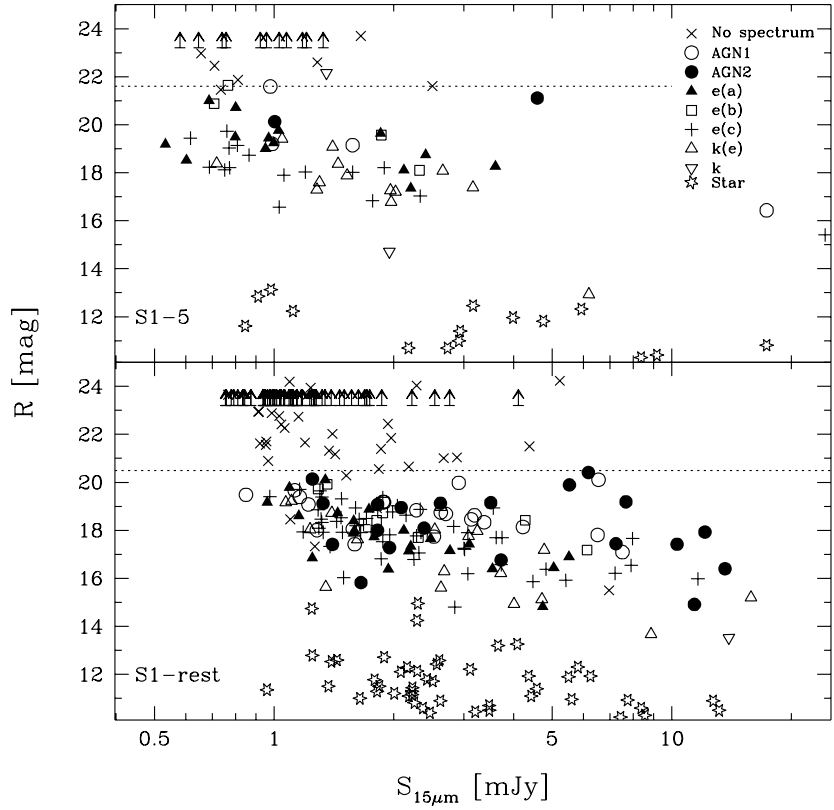}
\caption{
ELAIS-S1 spectroscopically identified sources in the S$_{15\mu
m}$--R-band plane. {\it Upper panel:} sources in the S1-5 area. {\it Lower
panel:} sources in the S1-rest area. The dashed lines are the
optical limits of the spectroscopically complete samples (see \S4.3), while
the upper arrows are sources without optical counterpart on CCDs.  }
\label{fig_SR}
\end{figure}

The S1\_5 area is 99$\%$ spectroscopically complete (72/73) down to
R=21.6. The S1\_rest area is 97$\%$ (216/222) complete down to R=20.5
(see Figures \ref{fig_histR} and \ref{fig_SR}). These magnitude limits
have been used for the statistical analysis of the evolution of AGNs
and galaxies by Matute et al. (in preparation) and Pozzi et al. (2004)
respectively. Both Figures \ref{fig_histR} and
\ref{fig_SR} show a separation between stars and galaxies, the stars
being optically brighter than galaxies, and also the appearance of a
separate population of galaxies at faint magnitudes ($R > 20$). See
\S6.3 for a detailed discussion on this point.

In the following we will call $noIDs$ all the 116 ELAIS-S1 sources
which are without spectroscopic identification (i.e., including also
72 EMPTY sources and 2 sources with no optical identification becouse
too close to bright stars; see \S3.2).


\section{Counts and contribution to the MIR background}

\begin{figure}
\centering
\resizebox{\hsize}{!}
{\includegraphics[angle=-90]{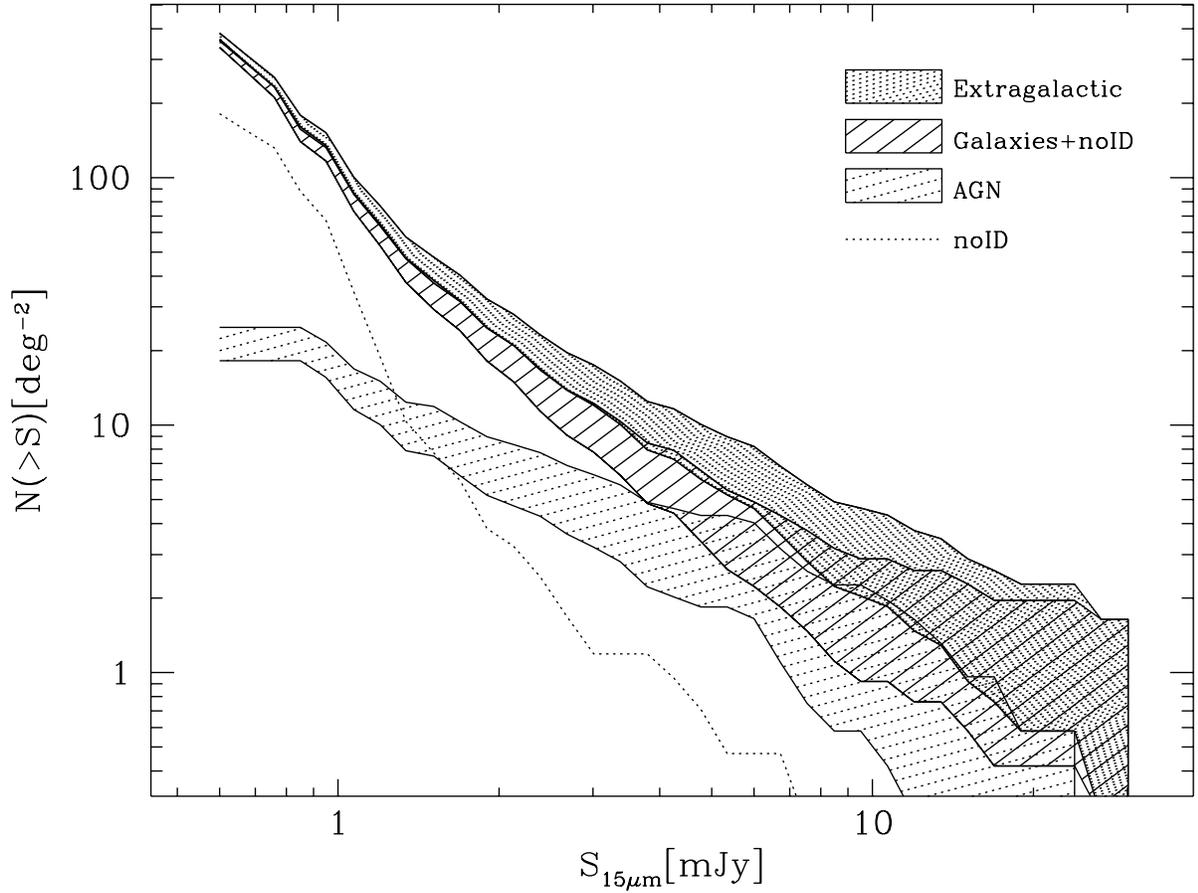}}
\caption{
Integral counts from the ELAIS-S1 survey for a) all the extragalactic
sources, b) the galaxies plus the no identified sources which are
assumed to be galaxies, c) the AGNs and d) the no identified sources. The counts
of galaxies and AGN corrected for the contribution of X-ray emitting AGNs
are also shown. The no identified sources ($noIDs$) includes 72+2 EMPTY
sources plus 40 (mainly faint: 20.5$<$R$<$23.3) sources without redshift.
Errors are at the 1$\sigma$ confidence level, based on Poisson
statistic.}
\label{fig_counts}
\end{figure}

We have computed the contribution of each class of sources to the 15
$\mu$m source counts in the mJy flux region.  The integral counts for
star-forming galaxies and AGNs (as defined on the basis of the optical
spectra) are reported in table \ref{tab_counts} (see Figure
\ref{fig_counts}). As already discussed in section \S4.2, the counts of
AGN should be considered as lower limits.
All ELAIS-S1 extragalactic sources (including the $noIDs$), with 15
$\mu$m fluxes brighter than 0.6 mJy, produce a 15 $\mu$m background of
$\nu$I($\nu$) = 0.34 nW m$^{-2}$sr$^{-1}$. This corresponds to
$\sim$12$\%$ of the total extragalactic background estimated by
Metcalfe et al. (2003) of $\nu$I($\nu$)=2.7 nW m$^{-2}$ sr$^{-1}$.

The sources spectroscopically identified with galaxies (no AGN)
produce a flux of $\nu$I($\nu$)=0.16 nW m$^{-2}$ sr$^{-1}$. The empty
fields (mainly objects fainter than R$\sim$23) produce
$\nu$I($\nu$)=0.14 nW m$^{-2}$ sr$^{-1}$, while the AGNs produce
$\nu$I($\nu$)=0.04 nW m$^{-2}$ sr$^{-1}$. If we assume that the $noID$
sources correspond mainly to starburst galaxies, we derive a
$\sim$90$\%$ contribution of starburst galaxies and a $\sim$10$\%$
contribution of AGNs to the 15 $\mu$m background at fluxes brighter
than 0.6 mJy. If our preliminary estimate of the fraction of of hidden
X-ray emitting AGN2 among the galaxies is taken into account, these
figures may well be consistent with the higher estimates
($\sim$15--20\%) for the contribution of AGN derived by Alexander
et~al. (2002), and Fadda et~al. (2002).


\section{Nature of the 15 $\micron$ extragalactic source population}

\subsection{Redshifts and luminosities}

\begin{figure}
\plotone{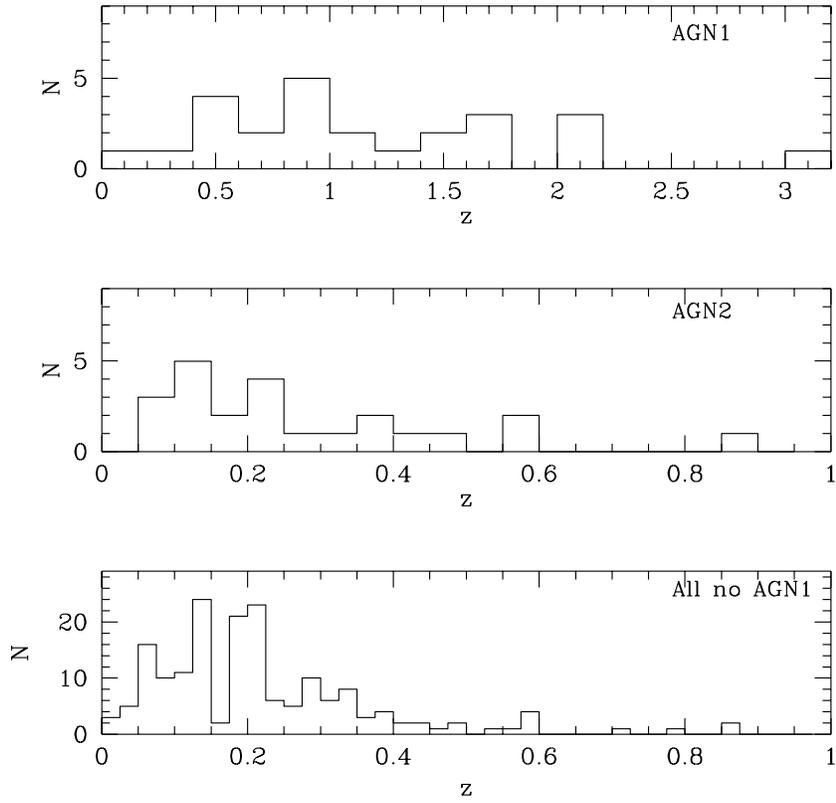}
\caption{
Redshift distributions of the spectroscopically identified sources
in ELAIS-S1. {\it Top}: AGN1; {\it Middle}: AGN2; {\it Bottom}: all the extragalactic
sources which are not AGN1. }
\label{fig_histz}
\end{figure}

In Figure \ref{fig_histz} the redshift distribution of the
spectroscopically identified sources is shown. While AGN1 reach
redshifts up to $z$$\sim$3, all AGN2 and galaxies have redshifts lower
than 1, with an average of $<$$z$$>$$\sim 0.3$. A possible hint of
structure at $z\sim0.15$ and $z\sim0.20$ is present.

\begin{figure}
\centering
\resizebox{\hsize}{!}
{\includegraphics[angle=-90]{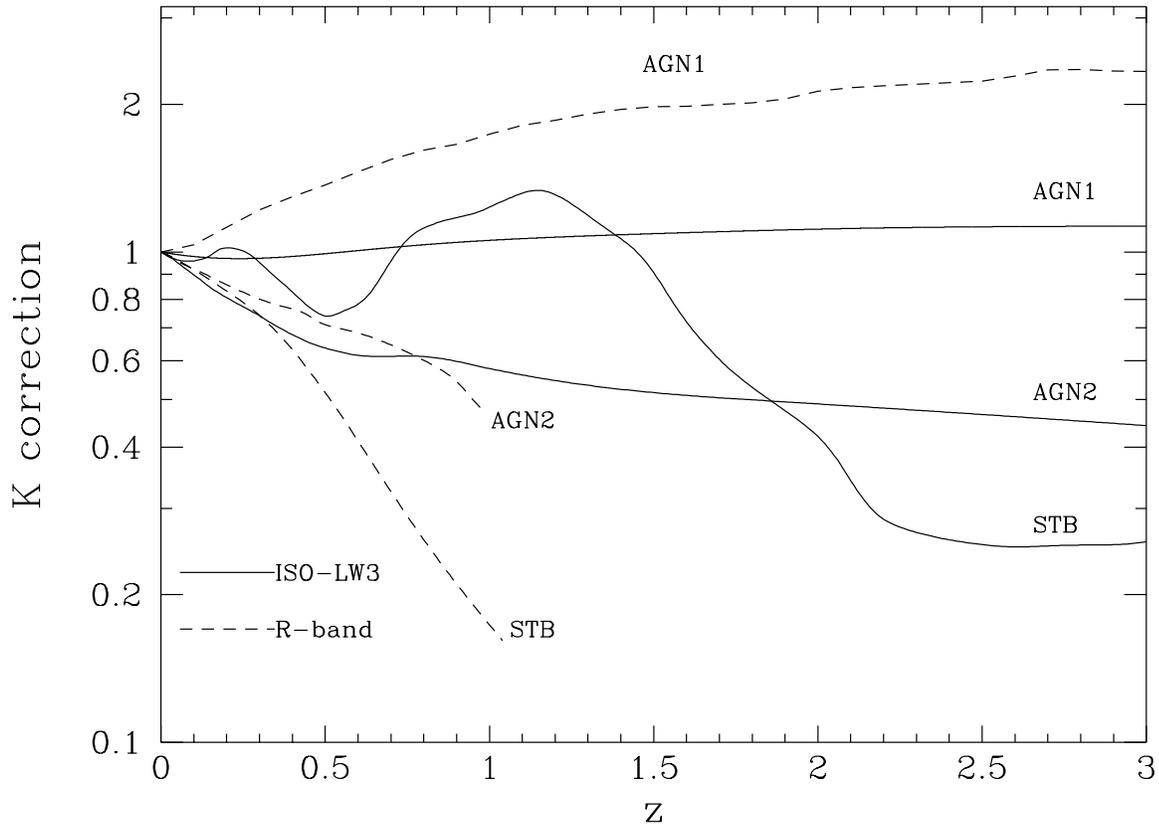}}
\caption{
K-correction ( log(S($z$)/S(0)) ) in the ISOCAM-LW3 15 $\micron$ and R-band filters for
type 1 AGN, type 2 AGN and Starburst galaxies. For Starburst galaxies
the MIR K-correction with M82 is shown as it is used in the vast majority
of the cases (see \S6.1).}
\label{fig_kcorr}
\end{figure}

We computed the $\nu$L$_\nu$ luminosities of our sources at 15
$\micron$ (L$_{15}$) and in the R-band (L$_{R}$). For each class of
objects a K-correction was applied according to the spectral
classification (see Figure \ref{fig_kcorr}). In the infrared the
K-corrections were based on: a) the SED from Elvis et al. (1994) for
AGN1, b) the SED of NGC1068 for AGN2 (Sturm et~al. 2000) and, c) the
SED of M82 for all of the remaining sources, except the absorption
line galaxies (for which M51 was considered). In the optical, the
K-correction from Natali et al. (1997) was used for AGN1, while for
AGN2 the K-corrections were computed by using the average of our
corresponding optical spectra.  For the galaxies the SEDs of M82 and
M51 were used, consistently with what has been done for the 15
$\micron$ flux. The model SEDs of both M82 and M51 were taken from the
GRASIL\footnote{available at
\url{http://web.pd.astro.it/granato/grasil/modlib/modlib.html}.} 
library of models (Silva et al. 1998). For M82, the MIR (5 -- 18
$\micron$) has been replaced with the observed ISOCAM CVF spectrum
(F\"orster Schreiber et al. 2003).



\begin{figure}
\centering
\resizebox{\hsize}{!}
{\includegraphics[angle=-90]{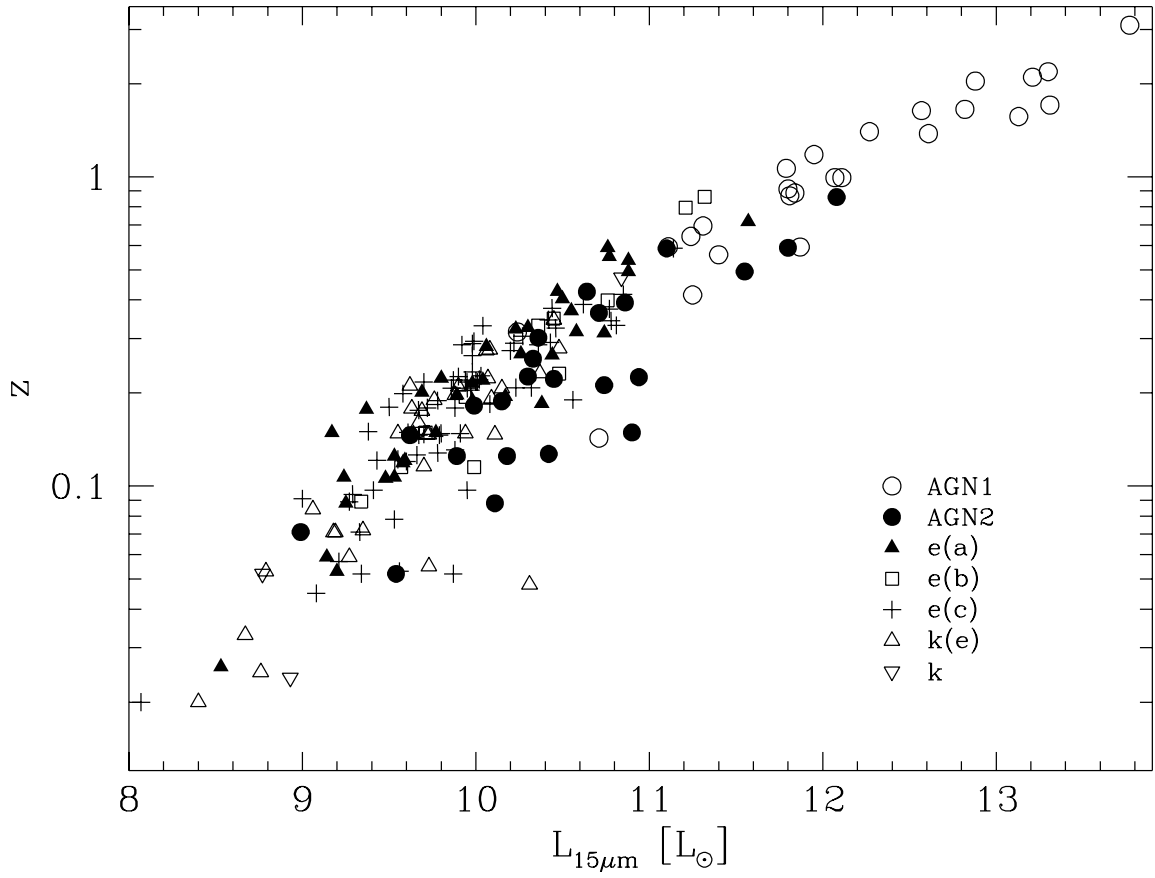}}
\caption{
Luminosity at 15 $\mu$m
of the ELAIS-S1 spectroscopically identified sources as a function of redshift.
}
\label{fig_zL}
\end{figure}

In Figure \ref{fig_zL} the $L_{15}$ luminosities are shown as a
function of redshift. A luminosity dependence of the classes of
objects is evident. AGN1 are the most luminous and most distant
sources with $<logL_{15}$$>$$\sim12.1$, while AGN2 are less
luminous and less distant with $<logL_{15}$$>$$\sim10.6$. 
Finally, the least luminous and least distant sources are
the galaxies with $<logL_{15}$$>$$\sim9.9$.

\subsection{The optical-MIR SED of starburst galaxies}

\begin{figure}
\centering
\resizebox{\hsize}{!}
{\includegraphics[angle=-90]{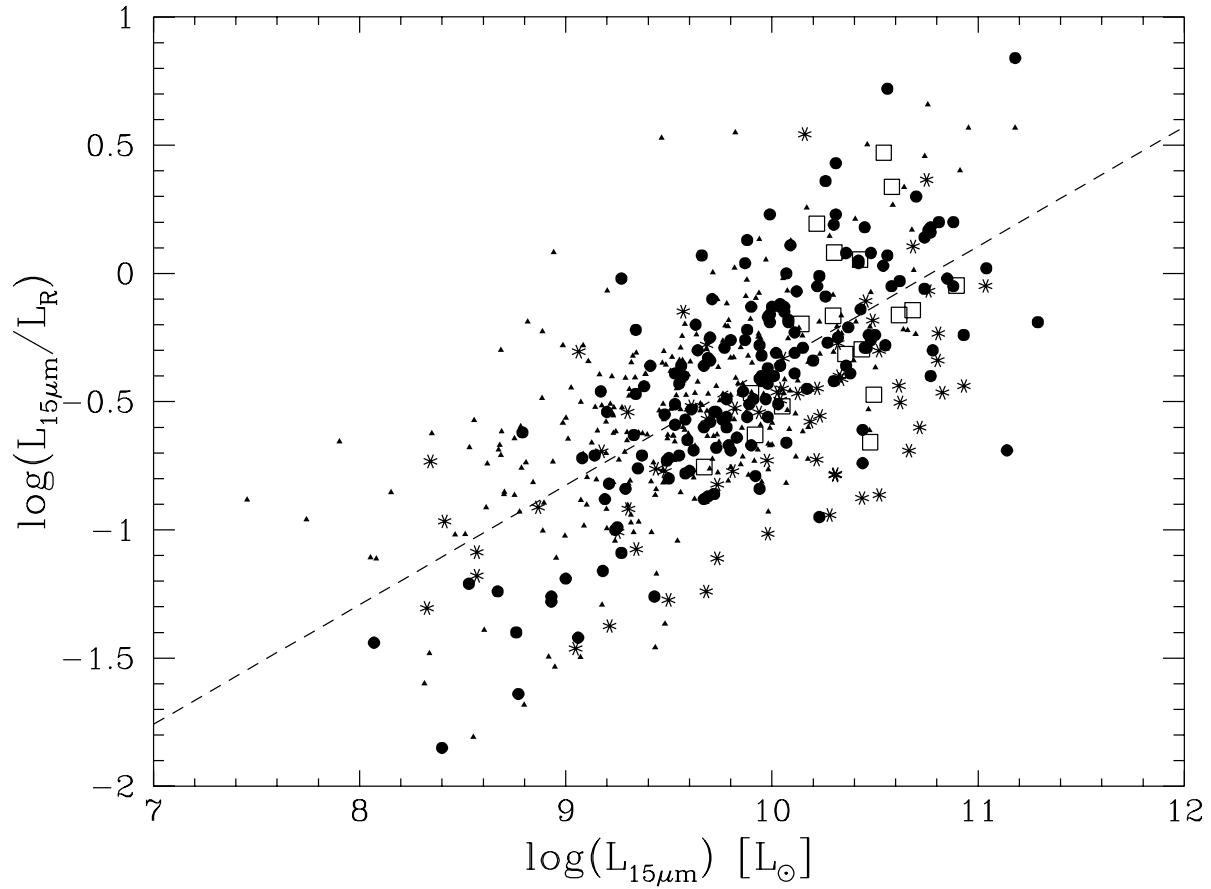}}
\caption{
L$_{15}/L_{R}$ ratio as a function of L$_{15}$ for different MIR
surveys. Filled circles are the ELAIS-S1 sample; open squares and
asterisks are data from HDF-S and HDF-N, respectively; filled triangles
are {\it IRAS} galaxies from Rush, Malkan and Spinoglio (1993).
Type 1 AGN have been excluded, due to the different mechanism responsible
for their luminosities.}
\label{fig_l15_lopt}
\end{figure}

In Figure \ref{fig_l15_lopt}, $L_{15}/L_{R}$ versus $L_{15}$ 
is shown for all of the ELAIS-S1 15 $\mu$m galaxies plus
sources from the spectroscopic identifications in the HDF-S from
Franceschini et al. (2003), the HDF-N (ISOCAM data from Aussel et
al. 1999; spectroscopic data from Cohen et al. 2000; magnitudes from
Hogg et al. 2000), and the local sample of {\it IRAS} 12-$\mu$m
galaxies of Rush, Malkan \& Spinoglio (1993; RMS; after conversion to
15 $\mu$m by using the M51 and M82 SED for normal and starburst
galaxies respectively). A clear
relation which is valid for all of the considered surveys is found,
although it involves only objects with optical identification. The
best fit to the data is $$ log(L_{15}/L_{R}) =
0.47logL_{15}[L_{\odot}]-5.02. $$

The 1$\sigma$ dispersion of the relation is 0.32. Note that type 1
AGN have been excluded because of the different mechanism responsible
for their luminosities, placing them on a different $L_{15}/L_{R}$
versus $L_{15}$ plane location with respect to galaxies. On the
contrary, type 2 AGN follow the same relation as all the other
galaxies.  This relation implies that the average optical-IR spectral
energy distribution of galaxies is luminosity dependent, with more
luminous galaxies having larger $L_{15}/L_{R}$ ratios. A two order of
magnitude increase in luminosity implies a 10 times larger
$L_{15}/L_{R}$ ratio. This relation is also discussed and taken into
account for the analyses of the evolution of starburst galaxies and
AGNs presented in two companion papers by Pozzi et al. (2004)
and Matute et al. (in preparation).


\subsection{The nature of the sources fainter than the spectroscopic or photometric limits}



As already shown by Gruppioni et al. (2002), the 15-$\micron$
extragalactic counts show an upturn at fluxes fainter than $\sim$1 my
(see also Elbaz et al. 1999), which is interpreted as due to the
contribution of the strong cosmological evolution of the starburst
population. This interpretation is discussed in detail by Pozzi et
al. (2004). Anyway the counts (see Figure
\ref{fig_counts}) show that the upturn is mostly due to the
contribution of the $noIDs$ sources (i.e. sources fainter than our
spectroscopic or photometric limits), which represent about half of the
extragalactic sources at fluxes $\sim$0.6-1 mJy.  These sources could
be either optically less luminous (absorbed?)  galaxies at redshift
similar to those of the already spectroscopically identified sample,
or higher redshift sources. We favor the second hypothesis on the
basis of the following discussion.

\begin{figure}
\centering
\epsscale{0.6}
\plotone{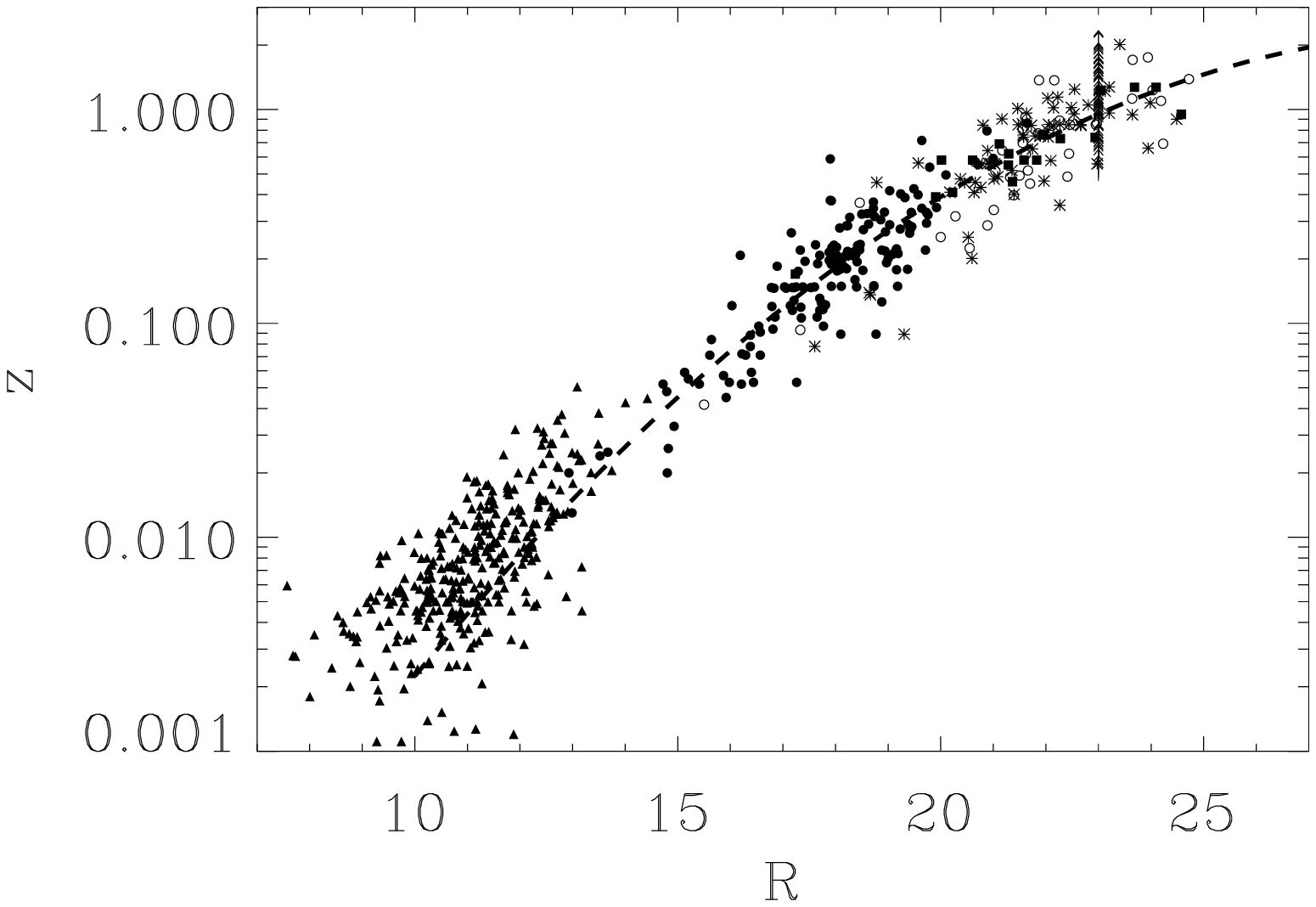}
\epsscale{0.55}
\plotone{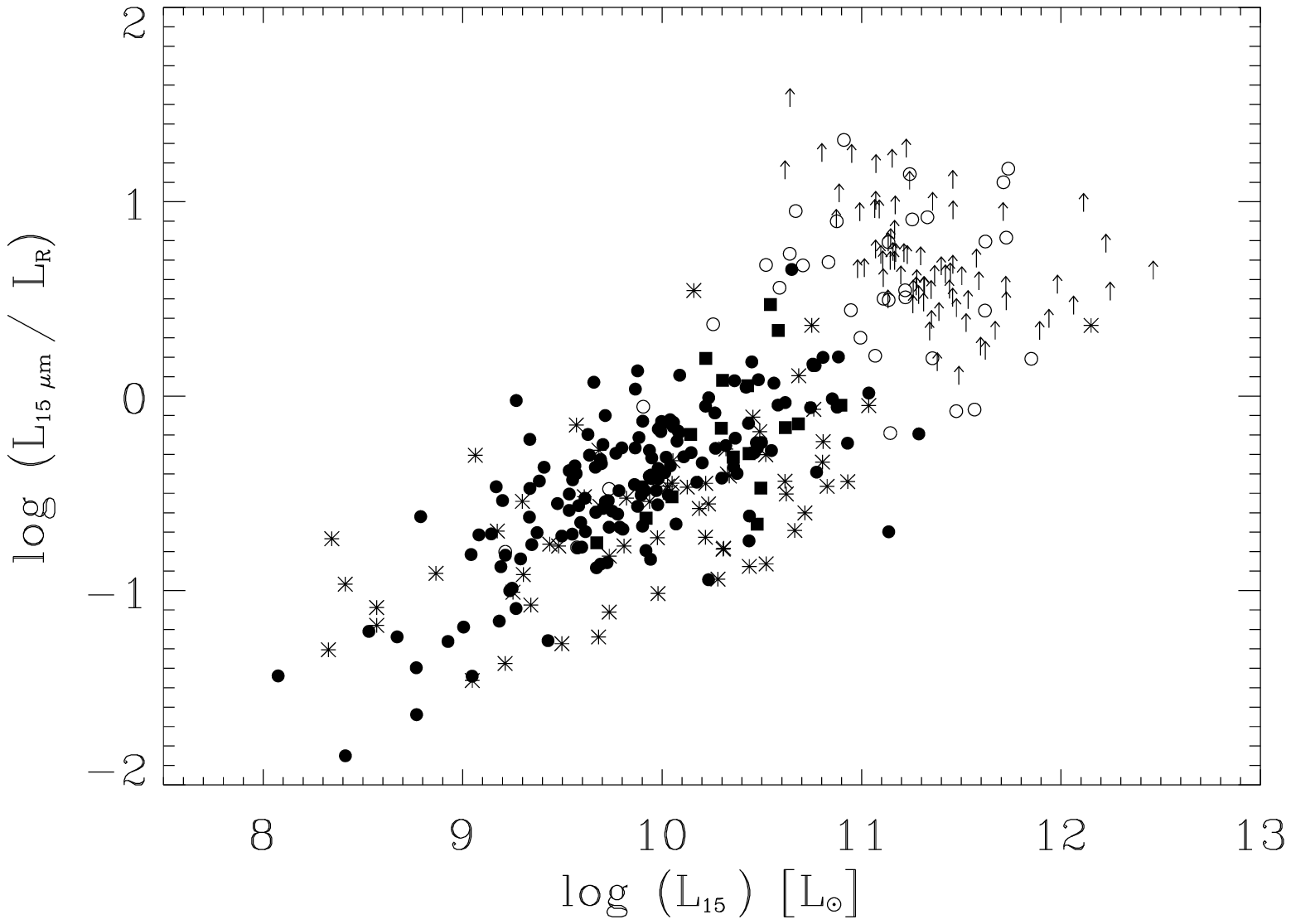}
\caption{
$Top$: $R-z$ plane with the lower limits estimates of the redshift for
the $noIDs$ sources as described in \S6.3. Filled circles are ELAIS-S1
data points; filled squares, asterisks and filled triangles are data
from HDF-S, HDF-N and RMS respectively. ELAIS-S1 sources with optical
counterparts, but lacking redshift information are shown as open
circles, while empty fields are represented as pointed-up arrows (all
at $R = 23$). The dashed line is the best-fitting polynomial function
used for estimating redshifts for sources without spectroscopic
identification.  $Bottom$: Same as in figure \ref{fig_l15_lopt}, but
with also the lower limits for the $noIDs$ sources shown. Symbols are
the same as in the $top$ panel.  Data from RMS are not shown in this
plot for greater clarity.}
\label{fig_fitz}
\end{figure}

We have tried to estimate a redshift for the $noIDs$ sources in order
to distinguish between the two different scenarios described above. We
have combined our sample with the fainter HDF-S and HDF-N
samples. Then we have tried to derive a flux-redshift relationship
applicable to the ELAIS-S1 unidentified sources in order to estimate
their redshifts. After several trials using multiple polynomial
relations between $log(z)$ and both 15 $\mu$m flux ($logS_{15}$) and
magnitude ($R$), we have assigned a redshift to the 116 $noIDS$
sources on the basis of their $R$ magnitudes only ($R=23$ was assumed
for all the unidentified 76 ($EMPTY$) sources), by considering the
following empirical relation and its spread:

$$ log(z) = -6.350 + 0.443~R  - 0.007~R^2 + G(0,\sigma_{rel}),$$
\noindent 
where $\sigma_{rel}$ (= 0.15) is the 1$\sigma$ dispersion of the
relation and $G(0,\sigma_{rel})$ is a Gaussian distribution with
center 0 and width $\sigma_{rel}$ (see Figure \ref{fig_fitz}, $Top$).
Note that the fit is non-linear in $log(z)$ vs. $R$, since we found
that a linear fit to the data extrapolated to faint magnitudes was
providing too high redshifts, inconsistent with the k-correction limit
of ISOCAM ($z \sim 1.5$) and with the median redshifts of $R = 24-25$
objects from the HDF-S and HDF-N.  Using the estimated redshifts, we
have computed MIR and optical luminosities for the unidentified
sources (by assuming all of them being star-forming galaxies, thus
using M82 SED) and we have placed also these objects in the $L_{15}/L_{R}$ vs.
$L_{15}$ diagram (see the $bottom$ panel of Figure \ref{fig_fitz}).

\begin{figure}
\plotone{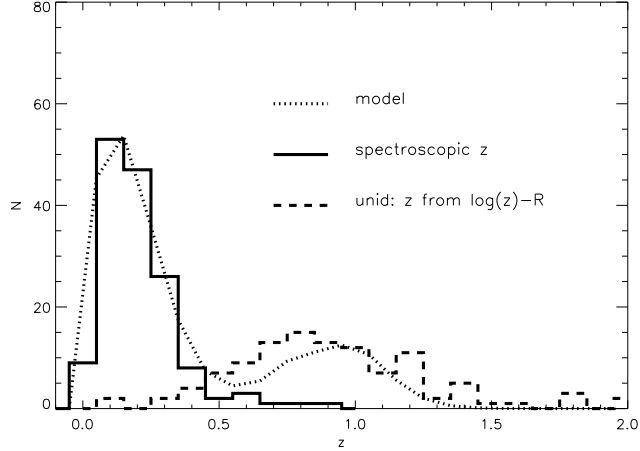}
\caption{
Measured redshift distribution for the spectroscopically identified
sources in S1 (solid histogram) and estimated redshift distribution
for the unidentified ones (dashed histogram). Data are compared with
the prediction of the model derived from the 15 $\mu$m luminosity
function obtained from our data and fitting the source counts (see
Pozzi et al. 2004), shown as dotted line.}
\label{fig_z_empty}
\end{figure}

According to our estimates, the unidentified objects in our survey are
likely to be rather luminous MIR objects ($L_{15}$ ranging
between $\sim 10^{11}$ L$_{\odot}$ and a few $10^{12}$ L$_{\odot}$, in the
LIG/ULIG region), and their optical and MIR estimated luminosities
appear to be consistent with (and extending to higher luminosities)
the $L_{15}/L_{R}$ vs. $L_{15}$ correlation found for the
identified sources. The estimated redshift distribution for the
unidentified sources is shown in Figure \ref{fig_z_empty}, where it is
compared with the distribution found for the spectroscopically
identified sources and with the model fitting derived from our data
(i.e. the 15 $\mu$m luminosity function in S1; Pozzi et al. 2004).
The agreement between the model and the data (either
observed and estimated) is good. It results that the unidentified
sources are expected to fill the secondary high-$z$ peak of the
model-predicted redshift distribution.

The double peaked redshift distribution (see Figure
\ref{fig_z_empty}) is caused by the 15 $\micron$ k-correction. As shown in
Figure \ref{fig_kcorr}, at $z$=0.5 the 15 $\micron$ observed fluxes
decrease by $\sim$30\%. For our flux limited sample this implies that
the ELAIS-S1 galaxies at $z$=0.5 correspond to 30\% higher intrinsic
luminosities.  But at $z$=0.5 the ELAIS-S1 galaxies sample the bright
steep slope of the luminosity function (log(L$_{15})$$>$11
$>$ L$^\ast$), where 30\% higher intrinsic luminosities
correspond to an order of magnitude lower density of galaxies. This
decrease of density of galaxies is also evident in the R-magnitude
distribution (see Figures \ref{fig_histR} and
\ref{fig_SR}), where, due to the k-correction effects, a lack 
of sources with R$\sim$20-21 (and $z$$\sim$0.5) is evident. 
The R-band histogram (Figure \ref{fig_histR}) shows a triple peaked
distribution: stars populate the brightest peak at
R$\sim$10-12, $z$$<$0.5 galaxies populate the intermediate peak at
R$\sim$17-19, and higher redshift ($z$$>$0.5) galaxies populate
the fainter peak that can be observed at R$>$21-22.

Under the alternative hypothesis that ELAIS-S1 $noIDs$ were instead lower
redshift ($z$$<$0.5) sources suffering from dust extinction, it would
be more appropriate to consider the 15-$\micron$ flux -- $z$ relation
rather than the $R-z$ relation (because affected by dust) to estimate
their redshifts. In this case, we would obtain a redshift distribution
for the $noIDs$ sources similar to that of the optically identified,
but they would be located in a separate region of the $L_{15}/L_{R}$
vs.  $L_{15}$ diagram (i.e. at similar $L_{15}$, but with at least one
order of magnitude higher $L_{15}/L_{R}$).  This would imply the
existence of a absolutely different population of objects, with much
higher level of obscuration (without continuity) than the identified
ones, but at similar flux density and redshifts. 


The first hypothesis is also supported by the faint spectroscopic
identifications in the HDF-N (Aussel et al. 1999) and CFRS 1452+52
(Flores et al. 1999) fields which have mainly found a population of
$z$=0.4-1.0 galaxies. For all these reasons we conclude that the vast
majority of the faint (R$>$20.5) ELAIS-S1 $noIDs$ objects should
belong to the same population as the identified sources, but at higher
redshifts ($z$=0.5-1.5), and that consequently the present sample is
virtually spectroscopically complete up to $z$=0.5, as far as the star
forming galaxies are concerned.

\subsection{The nature of the MIR galaxies}

As we have seen, our sample is virtually complete up to $z\simeq
0.5$. Up to this redshift we have identified 167 extragalactic
objects: 3 AGN1, 20 AGN2, 33 e(a), 9 e(b), 67 e(c), 32 k(e), and 3 k
galaxies. No k+a/a+k galaxy was found. The average redshift of the
sample is $<$$z$$>$=0.20.  The number of e(a) galaxies rises to 37 if
we classify also the AGN2 according to the classification scheme of
Dressler et al. (1999) (see \S4.2). The majority of the sources with
$z$$<$0.5 is composed by normal star forming spirals (e(c), 41\%), and
by absorbed star forming galaxies (e(a), 20\%; 23\% including the
AGN2). The total fraction of optical spectra which show emission lines
which are indicators of star formation (e(b)+e(c)+e(a)+k(e)) is 84\%.
14\% of the sources are AGNs, and only 2\% of the sources do not show
emission lines (k galaxies). This confirms that the MIR and the
optical data agree on revealing both ongoing star formation and
accretion driven (AGN) activities.  The lack of k+a/a+k galaxies in
the ELAIS-S1 sample suggest that these galaxies are not strong MIR
emitters. This result is consistent with the lack of emission lines in
their optical spectrum and with their post-starburst interpretation
(see Poggianti and Wu 2002).

The e(a) galaxies are believed to be associated with dust enshrouded
starburst galaxies.  Following Poggianti et al. (1999) and Poggianti
\& Wu (2000), the spectra of e(a) galaxies are explained as the result
of selective dust extinction which affects the youngest, most massive,
stars in HII regions much more than the older stellar population
responsible for the continuum flux. A search for similar spectra in
the local universe (Poggianti et al. 1999) revealed that they are
frequent among merging/strongly interacting galaxies ($\sim$40\%),
while they are scarce in optically selected surveys of field galaxies
at low $z$: 7-8\% in the Las Campanas Redshift Survey and in Kennicut
(1992). Dressler et al.  (1999) and Poggianti et al. (1999) found that
e(a) galaxies constitute about 10\% of their sample of cluster and
field galaxies at $z$$\sim$0.4-0.5.

At 15 $\mu$m most (71\%) of the sources with optical spectroscopy
in the CFRS are classified as e(a) galaxies and have a mean redshift
$<$$z$$>$=0.76, while at $z$=0.18, in the cluster of galaxies Abell 1869,
Duc et al. (2002) find 25\% of 15 $\mu$m galaxies classified as e(a).

The observed 20-23\% fraction of e(a) galaxies in the ELAIS-S1 sample
at $<$$z$$>$=0.20, is thus higher than observed in optically selected
samples and in agreement with the estimates at similar redshifts by
Duc et al. (2002) in a MIR selected sample. This confirms that e(a)
galaxies are dusty absorbed star forming galaxies which are
preferentially selected at longer wavelengths (MIR), and whose frequency
increases with increasing redshift, as the cosmic star formation rate.





\begin{figure}
\plotone{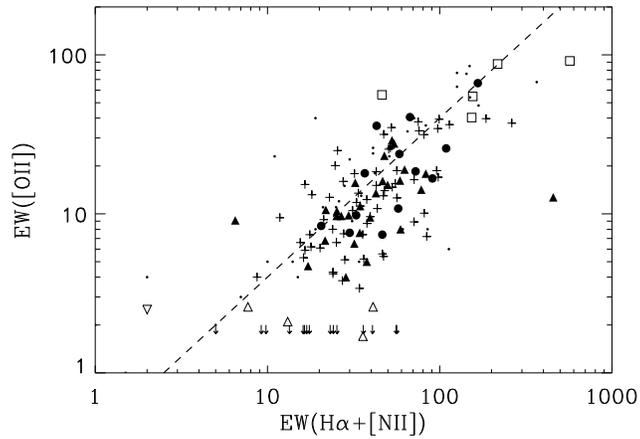}
\caption{
EW of [OII] versus EW of H$\alpha$+[NII] 
for ELAIS-S1 sources (symbols as in Figure \ref{fig_R15_RK}), and 39 {\it IRAS} galaxies from Rush,
Malkan \& Spinoglio (1993, small dots), using data from literature. The pointed-down arrows 
are sources with H$\alpha$+[NII] measurement but no [OII] detected. The dashed line shows
the relation $EW([OII]) = 0.4 EW(H\alpha+[NII])$ found by Kennicutt
(1992) for local field galaxies (see \S6.4).}
\label{fig_alphaoii}
\end{figure}

We have compared the $R-K$ colors of our sources to those of the field
galaxies and we did not find evidences of reddening. Becouse of the
bright K-band limit magnitude we limited our comparison in the
magnitude range R=15-17.5, and consequently the statistic are poor.
The average R-K color is 2.04 and 2.27 for ELAIS and field galaxies
respectively, with a 1$\sigma$ dispersion of $\sim$0.50 for both
samples. On the contrary, dust extinction seems to affect the spectral
lines of our objects. In figure \ref{fig_alphaoii} the EW of the [OII]
line is compared to the EW of the H$\alpha$+[NII] line.  The
$EW([OII])/EW(H\alpha+[NII])$ ratio is related to the differential
extinction due to dust, the $[OII]$ emission being more affected than
$H\alpha$ emission due to its shorter wavelength. The dashed line in
Figure \ref{fig_alphaoii} is the relation found by Kennicutt (1992)
for local field galaxies ($EW([OII])/EW(H\alpha+[NII])$$=$0.4).  Most
of our sources lie below that relation, and the presence of upper
limits (i.e. [OII] non-detections) would even lower the average value
of the $EW([OII])/EW(H\alpha+[NII])$, implying a larger amount of
extinction in our data than in the local sample of Kennicutt
(1992). The average ratio for the galaxies having both [OII] and
H$\alpha$+[NII] lines is 0.36$\pm$0.15, 0.35$\pm$0.03, 0.29$\pm$0.02
for e(b), e(c) and e(a) galaxies respectively. Thus, as expected, at
least for the normal star forming (e(c)) and absorbed star forming
(e(a)) galaxies the difference from the local observed ratio (0.4) is
significant.

\begin{figure}
\epsscale{0.5}
\plotone{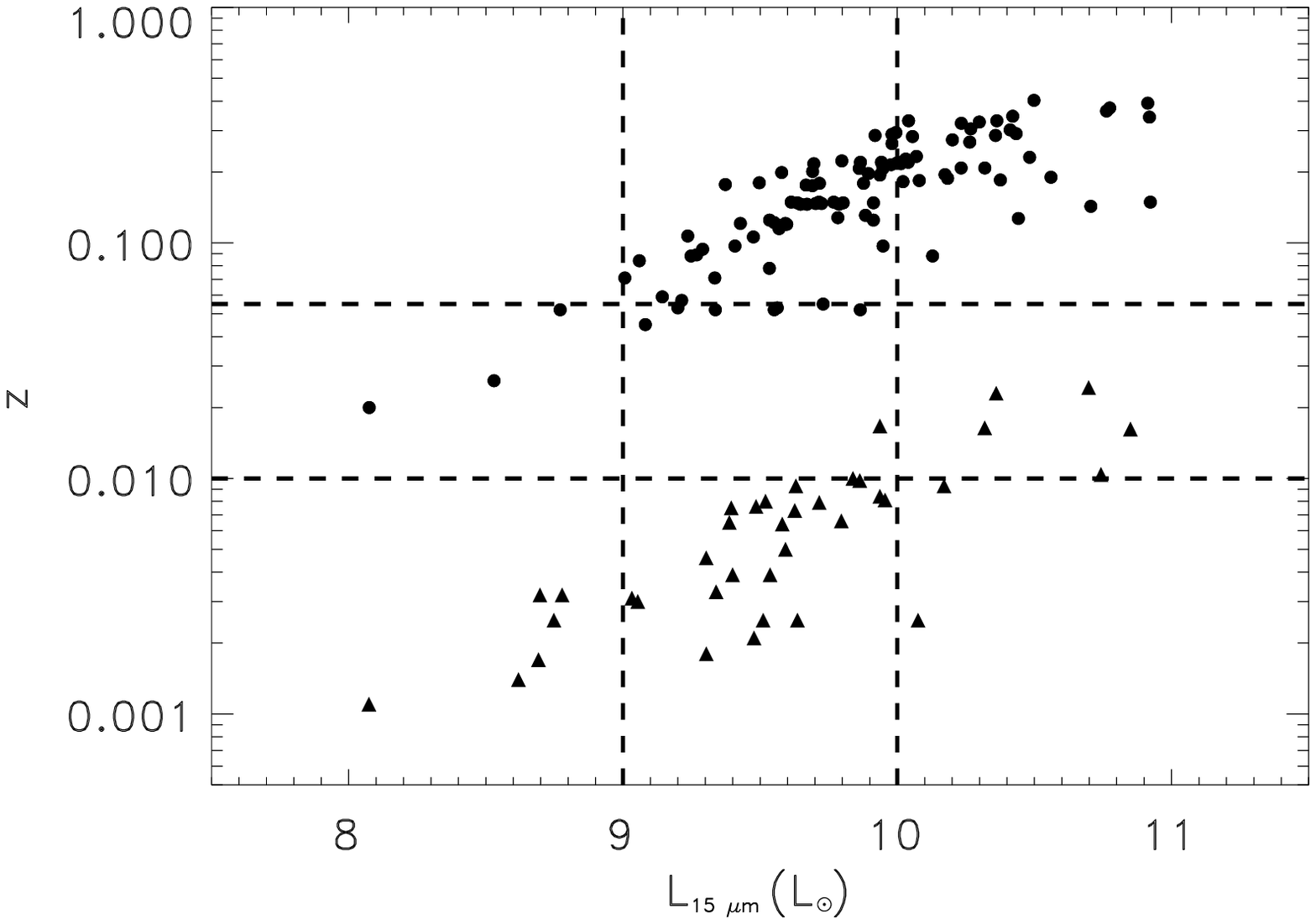}
\plotone{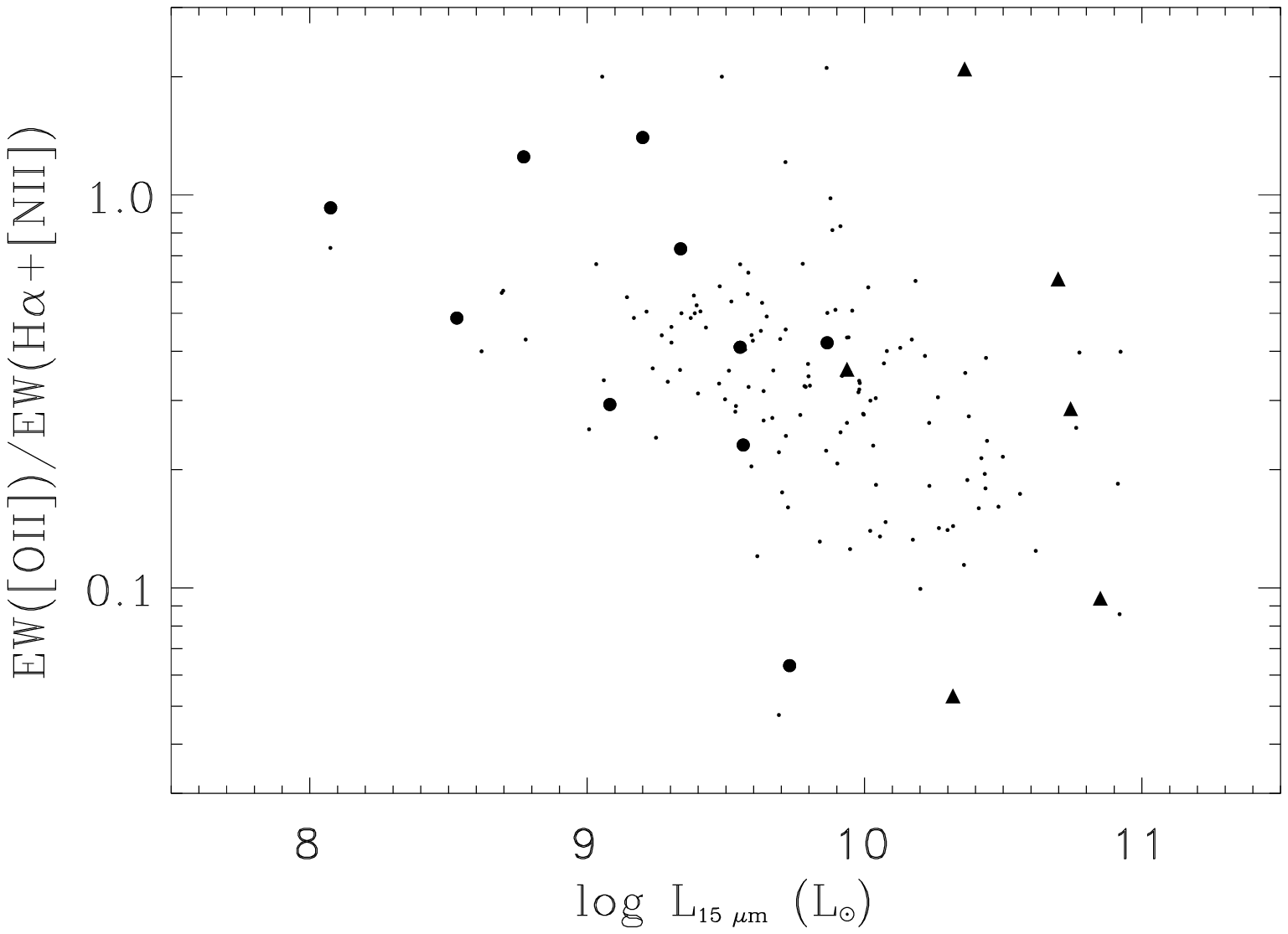}
\plotone{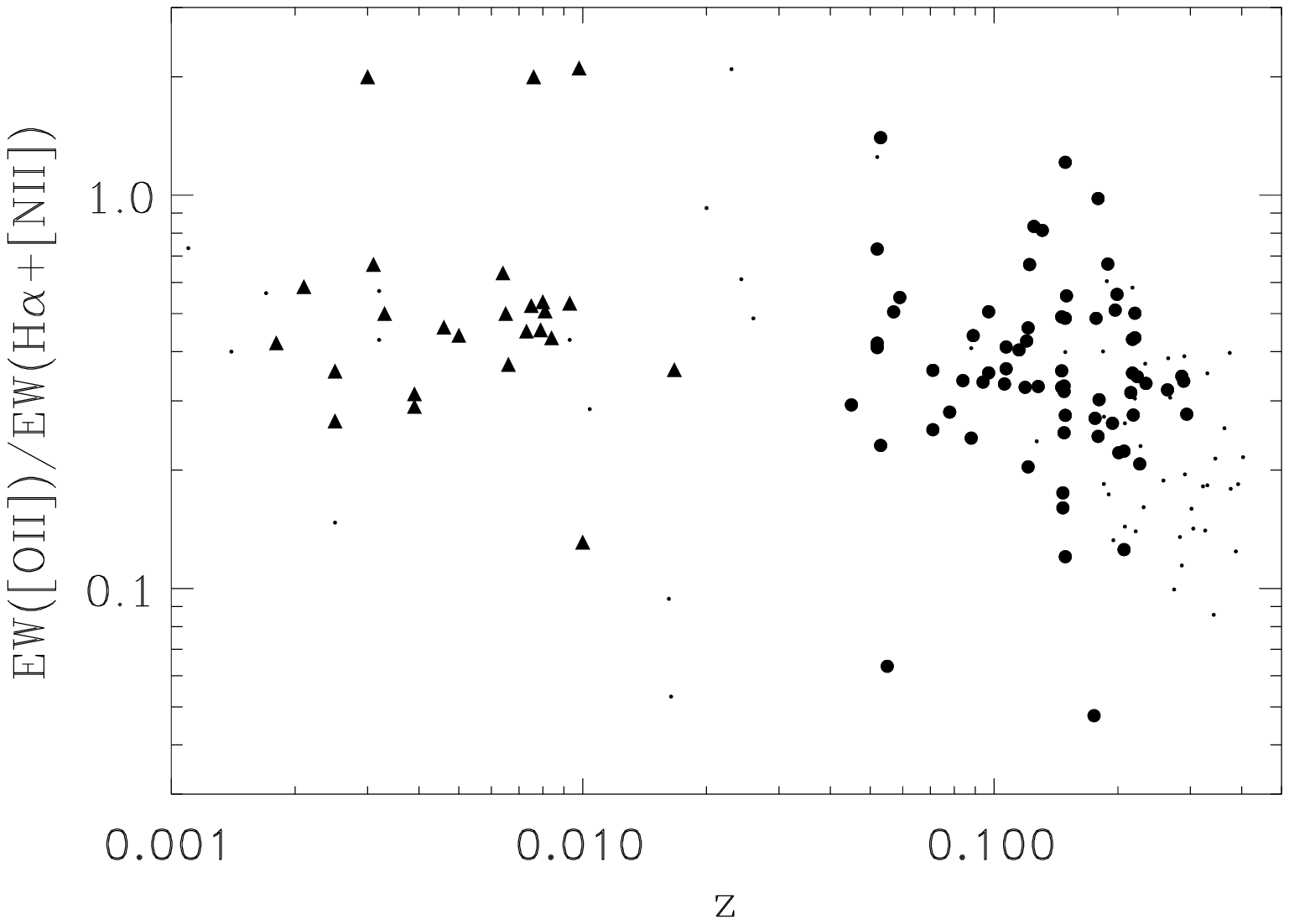}
\caption{$Top$ - Luminosity-redshift distribution of the ELAIS-S1 galaxies
(filled circles) plus 39 {\it IRAS} galaxies from RMS (triangles). Dashed lines show
the regions selected for disentangling the luminosity/redshift
dependence of the reddening (see \S6.4).
$Middle$ - 
$EW([OII])/EW(H\alpha+[NII])$ ratio versus 15 $\mu$m
luminosity for the galaxies with 0.010$<$$z$$<$0.055 (symbols as above).
$Bottom$ - 
$EW([OII])/EW(H\alpha+[NII])$ ratio versus redshift for the
galaxies with $9<logL_{15}(L_\odot)<10$ (symbols as above). The
remaining ELAIS-S1 plus RMS galaxies are represented with small
filled circles. 
\label{fig_linetrend}
}
\end{figure}

In order to disentangle a possible dependence on redshift or
luminosity of the average extinction of the galaxies, we have combined
our sample with a subsample of 39 galaxies from the local sample of
RMS for which EW measures for [OII] and H$\alpha$+[NII] were available
in the literature (see Alexander \& Aussel (2000) for a discussion of
this sample). The data show that the extinction affecting the lines is
an increasing function of luminosity while a correlation with redshift
is not statistically significant: a partial correlation Kendall test
gives in fact a 10$^{-4}$ probability that the EW is not dependent on
luminosity, and a 0.04 probability that the EW is not dependent on
redshift. Figure \ref{fig_linetrend} ($top$) shows the distribution of
the RMS and ELAIS galaxies in the luminosity-redshift plane. We have
then selected the galaxies with $0.010$$<$$z$$<$$0.055$ in order to
show the dependence on luminosity (Figure \ref{fig_linetrend},
$middle$), and the galaxies in a small luminosity interval
($9$$<$$logL_{15}(L_\odot)$$<$$10$) in order to show the not significant trend
with the redshift (Figure \ref{fig_linetrend}, $bottom$). These
results suggest an increase of dust extinction with luminosity.
Similar results, based on a smaller sample, were found also by Pozzi
et al. (2003) in one of the smaller but deeper ELAIS fields, S2.


A quantitative estimate of the amount of optical extinction affecting
the ELAIS-S1 MIR galaxy sample can be obtained by the measure of the
observed Balmer decrement in the H$\alpha$ and H$\beta$ lines. We have
made this measure on two composite spectra for each class of galaxies:
one including only spectra with detected H$\alpha$ and H$\beta$
emission lines, and one including also spectra with no detected
H$\beta$ line. Spectra showing H$\beta$ lines in absorption were
excluded.  This method was adopted in order to minimize the effects of
underling H$\beta$ absorption which causes the observed
H$\alpha$/H$\beta$ ratio to be larger. For this reason the extinction
estimates we derive have to be considered upper limits. In Table
\ref{tab_ext} our estimates in the case of B recombination with a
density of 100 cm$^{-3}$ and a temperature of 10000$^o$K (Osterbrock
1989) are shown. The higher values of E(B-V) were always obtained when
the spectra without H$\beta$ detection were included in the composite
spectra. These values correspond to an extinction
$A_V$=3.2$\times$$E(B-V)$ of about 1.5, 3.5, 4.0, 4.5 mag for e(b),
e(c), e(a) and k(e) galaxies respectively. These value refers to the
absorption of the lines emission; typically the absorption on the
continuum is a factor of two smaller (see also Pozzi et
al. 2003, and Poggianti \& Wu 2000 for similar results). As expected,
the larger extinction values are found for e(a) and k(e) galaxies,
reinforcing the interpretation that the e(a) signature is associated
with dust extincted starbursts.

\subsection{The star formation rate}

We have estimated the star formation rate (SFR) of the ELAIS-S1
sources from their 15 $\micron$ infrared luminosity. The SFR estimator
has been computed as: 

$$ SFR[M_{\odot}/yr]= 11.25\times10^{-23}L_{15}[W Hz^{-1}], $$

assuming a Salpeter initial mass function (IMF, Salpeter 1955) between
0.1 and 100 $M_{\odot}$, a dependence of SFR from L$_{60\mu m}$ as
given by Mann et al. (2002; $SFR[M_{\odot}/yr]=
0.9\times10^{-23}L_{60\mu m}[W Hz^{-1}]$), and a mean value of the
ratio $<L_{15 \mu m}[W Hz^{-1}] / L_{60\mu m}[W Hz^{-1}]>
\simeq 0.08$, typical of starburst galaxies such as M82.

\begin{figure}
\centering
\resizebox{\hsize}{!}
{\includegraphics[angle=-90]{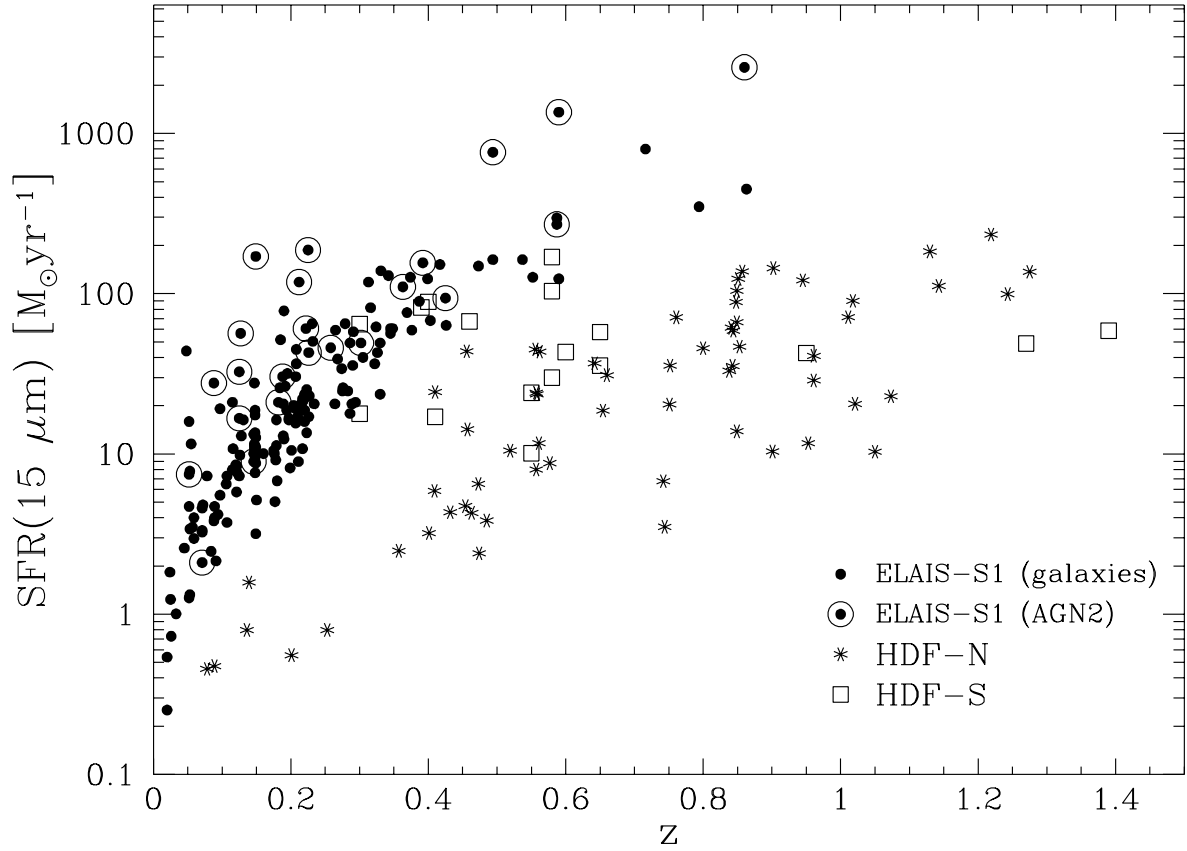}}
\caption{
Rates of the star formation based on the estimates of the MIR (15 
$\mu$m) luminosity, as a function of redshift.
Type 1 AGN have been excluded as their MIR
luminosity is not attributed to the star forming activity.}
\label{fig_zsfr}
\end{figure}

In Figure \ref{fig_zsfr} the SFRs as a function of redshift are
shown. Type 1 AGN have been excluded as their MIR luminosities are not
attributed to star forming activity, but to reprocessed higher
frequency luminosity driven by the accretion on a super-massive black
hole (10$^8$-10$^{10}$ M$_{\odot}$).
For the sake of comparison the
estimates from the spectroscopic identifications in the HDF-S, and the
HDF-N are shown. The ISOCAM observations of the HDF-S and HDF-N reach
15 $\micron$ fluxes down to $\sim$0.1 mJy (i.e.,\ more than six times
fainter than our survey) and probe a different region of the SFR
vs. $z$ plane. The shallower wide-area ELAIS survey is complementary
to the deep surveys and is required to make a more accurate estimate
of the SFR history of the Universe (see Pozzi et al. 2004).  The
ELAIS-S1 sources have mean SFRs of 13 M$_\odot$/yr, 45 M$_\odot$/yr and
242 M$_\odot$/yr, in the redshift ranges $z$$<$0.2, 0.2$<$$z$$<$0.4
and $z$$>$0.4 respectively.  Obviously this should not be attributed
to an evolutionary trend, but to the luminosity-redshift relation
which is present in any flux limited sample like ELAIS-S1.


We have found a good correlation between the SFRs derived from the 15
$\micron$ luminosity and those derived from the 1.4-GHz luminosity
(computed as $SFR[{M_{\odot}\over yr}]= 1.1\times10^{-21}({\nu\over
GHz})^{\alpha}L_{1.4GHz}[{W \over Hz}]$, assuming a Salpeter IMF
between 0.1 and 100 $M_{\odot}$, and a power-law spectrum with a
spectral index $\alpha\sim$0.7; $S_{\nu}{\propto}{\nu}^{-\alpha}$).
The correlation between the SFRs obtained through the two different
indicators reflects the radio-MIR luminosity correlation found for
ELAIS-S1 galaxies by Gruppioni et al. (2003; see also Elbaz et
al. 2002).

\section{Summary}

We have presented multiband (R, K, 1.4 GHz) and optical spectroscopic
identifications of a reliable subsample of 406 15-$\micron$ sources
from the ELAIS-S1 survey over the flux density range
0.5$<S_{15\micron}<$150 mJy. These observations substantially
contribute to the understanding of the evolution of star-forming
galaxies and AGN in the MIR, which is presented in two companion
papers by Pozzi et al. (2004) and Matute et al. (in preparation)
respectively. This paper has described the details of the ground-based
follow-up observations and data reduction, and has presented optical
identifications for $\sim$82$\%$ of the 15 $\micron$ sources down to
R$\sim$23, and spectroscopic identification for $\sim$88$\%$ of the
optically identified sample. The areal coverage and completeness of
the survey has been described in detail, in order to allow statistical
analyses. The source counts for star-forming galaxies and AGN have
been derived: the ELAIS extragalactic sources down to
S$_{15\micron}$=0.6 mJy have a density of 370$\pm$14/deg$^2$, and
produce a 15 $\micron$ background of $\nu$I($\nu$) = 0.34 nW
m$^{-2}$sr$^{-1}$. This corresponds to $\sim$12$\%$ of the total
extragalactic background estimated by Metcalfe et al. (2003).  Taking
into account the optical based classification, we estimate an AGN
contribution of at least $\sim$6\% to the MIR sources brighter than
S$_{15\micron}$=0.6 mJy. A larger fraction is expected if the
classification will be based on the X-ray luminosity.  The line ratios
of the optical spectra of the MIR galaxies suggest that they are
affected by substantial reddening, whith the more MIR-luminous sources
having larger dust extinction. A general trend is found in the
MIR-optical SEDs of {\it IRAS} and {\it ISO} galaxies whith the more
MIR-luminous sources having larger $L_{15}/L_{R}$ ratios. According to
our analyses we conclude that most of the ELAIS galaxies fainter than
R$\sim$23 should have redshift greater than 0.5 and are likely to be
LIG-ULIG galaxies. As a consequence, the present sample of
star-forming galaxies has to be considered virtually complete up to
$z=0.5$.



\acknowledgments
Based on observations collected at the European Southern Observatory,
Chile, ESO N$^{\circ}$: 57.A-0752, 58.B-0511, 59.B-0423, 61.B-0146,
62.P-0457, 67.A-0092(A), 68.A-0259(A), 69.A-0538(A), 70.A-0362(A);
PATT proposal P076). This paper is based on observations with the
Infrared Space Observatory (ISO). ISO is ane ESA project with
instruments funded by ESA Member States (especially the PI countries:
France, Germany, the Netherlands and the United Kingdom) and with the
partecipation of ISA and NASA. This research has made use of the
NASA/IPAC Extragalactic Database (NED) which is operated by the Jet
Propulsion Laboratory, California Institute of Technology, under
contract with the National Aeronautics and Space Administration. This
publication makes use of data products from the Two Micron All Sky
Survey, which is a joint project of the University of Massachusetts
and the Infrared Processing and Analysis Center/California Institute
of Technology, funded by the National Aeronautics and Space
Administration and the National Science Foundation.  This research has
been partially supported by ASI contracts ARS-99-75, ASI I/R/107/00,
ASI I/R/113/01, MURST grants Cofin-98-02-32, Cofin-00-02-36. IM
acknowledges a PhD grant from CNAA/INAF. DMA acknowledges the Royal
Society for generous support. Ph. H\'eraudeau acknowledges the
financial support provided through the European Community's Human
Potential Programme under contract HPRN-CT-2002-00316, SISCO.

 We wish to thank Lucia Pozzetti for interesting discussions,
and the anonymous referee for the useful comments.

\clearpage

%

\clearpage

\clearpage



\clearpage

\begin{deluxetable}{ccc}
\tabletypesize{\tiny}
\tablewidth{0pt}
\tablecaption{The completeness function of S1-5 and S1-rest}
\tablehead{
\colhead{Flux }&  
\colhead{S1-5 }&  
\colhead{S1-rest } 
}
\startdata
   0.45&    0.001  &  0.000  \\
   0.50&    0.007  &  0.000  \\
   0.55&    0.023  &  0.000  \\
   0.60&    0.053  &  0.000  \\
   0.65&    0.110  &  0.001  \\
   0.70&    0.167  &  0.004  \\
   0.75&    0.239  &  0.009  \\
   0.80&    0.312  &  0.019  \\
   0.85&    0.379  &  0.031  \\
   0.90&    0.442  &  0.049  \\
   0.95&    0.501  &  0.076  \\
   1.00&    0.545  &  0.107  \\
   1.05&    0.585  &  0.147  \\
   1.10&    0.625  &  0.189  \\
   1.15&    0.668  &  0.232  \\
   1.20&    0.708  &  0.278  \\
   1.25&    0.743  &  0.323  \\
   1.30&    0.776  &  0.373  \\
   1.35&    0.807  &  0.415  \\
   1.40&    0.841  &  0.465  \\
   1.45&    0.866  &  0.506  \\
   1.50&    0.889  &  0.559  \\
   1.60&    0.929  &  0.646  \\
   1.70&    0.957  &  0.699  \\
   1.80&    0.977  &  0.743  \\
   1.90&    0.991  &  0.781 \\
   2.00&    1.000  &  0.813\\
   2.10&    1.000  &  0.841\\
   2.20&    1.000  &  0.867\\
   2.30&    1.000  &  0.888\\
   2.40&    1.000  &  0.906\\
   2.50&    1.000  &  0.921\\
   2.60&    1.000  &  0.934\\
   2.70&    1.000  &  0.945\\
   2.80&    1.000  &  0.954\\
   2.90&    1.000  &  0.962\\
   3.00&    1.000  &  0.969\\
   3.10&    1.000  &  0.974\\
   3.20&    1.000  &  0.979\\
   3.30&    1.000  &  0.983\\
   3.40&    1.000  &  0.986\\
   3.50&    1.000  &  0.989\\
   3.60&    1.000  &  0.991\\
   3.70&    1.000  &  0.993\\
   3.80&    1.000  &  0.995\\
   3.90&    1.000  &  0.996\\
   4.00&    1.000  &  0.997\\
   4.10&    1.000  &  0.999\\
   4.20&    1.000  &  1.000\\
\enddata
\label{tab_compl}
\end{deluxetable}

\begin{deluxetable}{ccrcccccrrrcrrrc}
\tabletypesize{\tiny}
\rotate
\tablewidth{0pt}
\tablecaption{Sources in the ELAIS S1 Sample}
\tablehead{
\colhead{ Name }&  
\colhead{ \# }& 
\colhead{ F$_{15}$ }& 
\colhead{ $\alpha_O$(J2000) } & 
\colhead{ $\delta_O$(J2000) } & 
\colhead{ Off } & 
\colhead{ R } & 
\colhead{ K } & 
\colhead{ Lik } & 
\colhead{ Rel } & 
\colhead{ F$_{1.4}$ } & 
\colhead{ $z$ }& 
\colhead{ Cl }& 
\colhead{ L$_{15}$ } & 
\colhead{ L$_R$ } & 
\colhead{ Notes }\\
\colhead{ (1) }&  
\colhead{ (2) }&  
\colhead{ (3) }&  
\colhead{ (4) }&  
\colhead{ (5) }&  
\colhead{ (6) }&  
\colhead{ (7) }&  
\colhead{ (8) }&  
\colhead{ (9) }&  
\colhead{ (10) }&  
\colhead{ (11) }&  
\colhead{ (12) }&  
\colhead{ (13) }&  
\colhead{ (14) }&  
\colhead{ (15) }&  
\colhead{ (16) }

} 
\startdata
ELAISC15\_J003957-432013&8&  3.01&00:39:57.85&-43:20:13.4& 1.2    &17.21&15.32&363.16    &0.999& $<$  0.24& 0.128&5\phm{$L$}& 9.78&10.27&  \\ 
ELAISC15\_J003958-441511&9& 25.39&00:39:58.67&-44:15:10.6& 2.0    &10.41& 7.39&999.99    &1.000& $<$  0.30& 0.000&8\phm{$L$}& \nodata& \nodata&  \\ 
ELAISC15\_J004000-431325&8&  1.01&\nodata&\nodata&\nodata    & \nodata&    \nodata&  \nodata    &\nodata& $ $  \nodata& \nodata&0\phm{$L$}& \nodata& \nodata&  \\ 
ELAISC15\_J004004-433710&9&  1.24&\nodata&\nodata&\nodata    & \nodata&    \nodata&  \nodata    &\nodata& $ $  \nodata& \nodata&0\phm{$L$}& \nodata& \nodata&  \\ 
ELAISC15\_J004007-432516&8&  3.11&00:40:06.88&-43:25:18.2& 2.5    &12.21& 9.46&999.99    &1.000& $<$  0.30& 0.000&8\phm{$L$}& \nodata& \nodata&  \\ 
ELAISC15\_J004009-434424&9&  2.62&00:40:09.26&-43:44:25.3& 0.9    &19.13&    \nodata& 97.33    &0.997& $<$  0.30& 0.188&2\phm{$L$}&10.15& 9.87&  \\ 
ELAISC15\_J004010-432012&8&  1.03&\nodata&\nodata&\nodata    & \nodata&    \nodata&  \nodata    &\nodata& $ $  \nodata& \nodata&0\phm{$L$}& \nodata& \nodata&  \\ 
\enddata
\tablecomments{ 
The complete version of this table is in the electronic edition of the
Journal.  The printed edition contains only a sample.  The columns are
as follows: (1) Name of the 15-$\mu$m source as in Lari et
al. (2001). (2) Raster number. (3) The 15-$\mu$m flux in mJy units.
(4-5) The ra and dec of the optical counterpart, J2000. (6) Offset of
the optical counterpart from the 15-$\mu$m source position in arcsec
units. (7-8) The R and K band Vega magnitudes. (9) Likelihood of the
optical identification; 999.99 was assigned to stars (see \S3.2). (10)
Reliability of the optical identification (see
\S3.2).  (11) 1.4 GHz flux or 3$\sigma$ upper limit in mJy units. (12)
Redshift. (13) Spectroscopic class where: 0 means no spectra
available, 1 is AGN1, 2 is AGN2, 3 is e(a) galaxy, 4 is e(b) galaxy, 5
is e(c) galaxy, 6 is k(e) galaxy, 7 is k galaxy and 8 is star. An $L$
is appended for objects classified LINERS according to the line ratios
diagnostics (see \S4.2).  (14) log($\nu L\nu$) 15-$\mu$m luminosity in
solar units.  (15) log($\nu L\nu$) R band luminosity in solar units.
(16) Notes, O is for outside CCD observations, M is for multiple
optical counterparts, and B for Broad Absorption Line (BAL) QSO.
}
\label{tab_id}
\end{deluxetable}

\begin{deluxetable}{crccrrrrrrrr}
\tabletypesize{\tiny}
\rotate
\tablewidth{0pt}
\tablecaption{Rest frame line measurements}
\tablehead{
\colhead{Name  }&  
\colhead{F$_{15}$  }&  
\colhead{$z$  }&  
\colhead{ Cl  }&  
\colhead{\phm{On}D$_{4000}$  }&  
\colhead{EW([OII])   }&  
\colhead{EW(H$\delta$)  }&  
\colhead{EW(H$\beta$)  }&  
\colhead{EW([OIII])  }&  
\colhead{EW(H$\alpha$)  }&  
\colhead{EW([NII])  }&  
\colhead{EW([SII])  }  \\
\colhead{ (1) }&  
\colhead{ (2) }&  
\colhead{ (3) }&  
\colhead{ (4) }&  
\colhead{ (5) }&  
\colhead{ (6) }&  
\colhead{ (7) }&  
\colhead{ (8) }&  
\colhead{ (9) }&  
\colhead{ (10) }&  
\colhead{ (11) }&  
\colhead{ (12) } 
}
\startdata
ELAISC15\_J002904-432415&  1.61&0.207&5\phm{$L$}&    1.20&   -8.9&     3.1&    -3.7&     0.0&   -56.7&   -10.6&   -12.0\\ 
ELAISC15\_J002915-430333&  2.04&0.417&5\phm{$L$}&    1.23&  -39.2&     3.4&   -10.3&   -13.9& \nodata& \nodata& \nodata\\ 
ELAISC15\_J002924-432233&  2.34&0.374&5$L$ &    1.17&  -39.4&     0.0&    -3.4&     0.0&   -36.4&   -47.3&   -46.3\\ 
ELAISC15\_J002939-430625&  3.73&0.071&5\phm{$L$}&    1.22&  -11.8&     3.6&    -1.9&     0.0&   -22.7&    -8.6&    -8.4\\ 
ELAISC15\_J002949-430703&  1.33&0.302&2\phm{$L$}&    1.40&   -7.4&     0.0&     0.0&   -25.0&   -15.8&   -23.8&   -12.9\\ 
ELAISC15\_J003001-432202&  1.47&0.274&5\phm{$L$}&    1.33&   -3.4&     3.4&     0.0&     0.0&   -28.1&    -8.0&     0.0\\ 
ELAISC15\_J003011-432947&  1.35&0.084&6\phm{$L$}&    1.87&   -2.6&     0.0&    -1.1&     0.0&    -4.8&    -3.4&     0.0\\ 
\enddata
\tablecomments{The complete version of this table is in the electronic edition of the
Journal.  The printed edition contains only a sample.  The columns are
as follows: (1) Name of the 15-$\mu$m source as in Lari et
al. (2001). (2) The 15-$\mu$m flux in mJy units.  (3) Redshift. (4)
Spectroscopic class as in Table \ref{tab_id}. (5) Calcium
break. (6-12) Rest frame EW in \AA\ units of the [OII]$\lambda$3727,
H$\delta$, H$\beta$, [OIII]$\lambda$5007, H$\alpha$,
[NII]$\lambda$6583, and [SII]$\lambda$6725 lines; positive values mean
absorption. }
\label{tab_lines}
\end{deluxetable}

\begin{deluxetable}{rcccccccc}
\tabletypesize{\tiny}
\tablecaption{The integral extragalactic counts at 15 $\micron$}
\tablewidth{0pt}
\tablehead{
\colhead{S$_{15}$    }&  
\colhead{All  }&  
\colhead{Galaxies\tablenotemark{a}  }&  
\colhead{AGN1+AGN2\tablenotemark{b}} &
\colhead{AGN1   } &
\colhead{AGN2\tablenotemark{b}  } \\
\colhead{(mJy)  }&  
\colhead{(N($>$S)/deg$^2$)  }&  
\colhead{(N($>$S)/deg$^2$)  }&  
\colhead{(N($>$S)/deg$^2$)  }&  
\colhead{(N($>$S)/deg$^2$)  }&  
\colhead{ (N($>$S)/deg$^2$) } 
}
\startdata
  0.60 &   370$\pm$14  & 349   $\pm$13   & 21.5$\pm$3.3 & 13.6 $\pm$2.6   &7.9 $\pm$2.0   \\
  0.76 &   242$\pm$11  & 221   $\pm$10   & 21.5$\pm$3.3 & 13.6 $\pm$2.6   &7.9 $\pm$2.0   \\
  0.95 & 140.8$\pm$8.4~& 122.2 $\pm$7.8  & 18.6$\pm$3.1 & 10.7 $\pm$2.3   &7.9 $\pm$2.0   \\ 
  1.20 &  70.1$\pm$5.9 &  57.6 $\pm$5.4  & 12.5$\pm$2.5 &  6.1 $\pm$1.7   &6.4 $\pm$1.8   \\
  1.51 &  42.4$\pm$4.6 &  32.7 $\pm$4.0  &  9.7$\pm$2.2 &  4.8 $\pm$1.6   &4.9 $\pm$1.6   \\
  1.90 &  22.8$\pm$3.7 &  20.7 $\pm$3.2  &  7.1$\pm$1.9 &  3.2 $\pm$1.3   &3.9 $\pm$1.4   \\
  2.39 &  19.3$\pm$3.1 &  13.3 $\pm$2.6  &  6.0$\pm$1.7 &  2.9 $\pm$1.2   &3.1 $\pm$1.2   \\
  3.01 &  14.1$\pm$2.7 &   9.3 $\pm$2.2  &  4.8$\pm$1.5 &  1.9 $\pm$1.0   &2.9 $\pm$1.2   \\
  3.79 &   9.5$\pm$2.2 &   5.9 $\pm$1.7  &  3.6$\pm$1.3 &  1.2 $\pm$0.8   &2.4 $\pm$1.1   \\
  4.77 &   7.3$\pm$1.9 &   4.3 $\pm$1.5  &  3.1$\pm$1.2 &  1.0 $\pm$0.7   &2.1 $\pm$1.0   \\
  6.00 &   5.7$\pm$1.7 &   2.8 $\pm$1.2  &  2.8$\pm$1.2 &  1.0 $\pm$0.7   &1.8 $\pm$1.0   \\
  7.55 &   3.6$\pm$1.3 &   1.90$\pm$0.97 &  1.7$\pm$0.9 & ...             &1.4 $\pm$0.8   \\
  9.51 &   2.6$\pm$1.1 &   1.19$\pm$0.77 &  1.4$\pm$0.8 & ...             &1.2 $\pm$0.8   \\
 11.97 &   1.9$\pm$1.0 &   0.95$\pm$0.69 &  1.0$\pm$0.7 & ...             &    ...        \\
 15.07 &  1.19$\pm$0.77&   0.71$\pm$0.60 &  ...         & ...             &    ...       \\
 18.98 &  0.71$\pm$0.60&   0.48$\pm$0.49 &  ...         & ...             &    ...       \\
\enddata
\tablenotetext{a}{
Including sources without spectroscopic classification ($noIDs$).}
\tablenotetext{b}{
These values should be considered lower limits (see \S4.2 and \S5).}
\label{tab_counts}
\end{deluxetable}

%
%

\begin{deluxetable}{lc}
\tablecaption{Mean E(B-V)}
\tablewidth{0pt}
\tablehead{
\colhead{Class    }&  
\colhead{E(B-V)    }
}
\startdata
e(b).....     &  0.43-0.52 \\
e(c).....     &  0.82-1.59 \\
e(a).....     &  1.02-1.42 \\
k(e).....     &  0.64-1.87 \\
\enddata
\label{tab_ext}
\end{deluxetable}


\end{document}